\documentclass{report}\newcommand{\emul}{true}

\newcommand{\IncludeFigures}{true}

\usepackage{amsmath}
\usepackage{amssymb}
\usepackage{xspace}

\usepackage{ifthen}
\newboolean{emul}%
\setboolean{emul}{\emul}%

\ifthenelse{\boolean{emul}}{
  \usepackage[longnamesfirst]{natbib}
  \bibpunct[]{(}{)}{;}{a}{}{,}
  \usepackage{emulateapj}
  \usepackage{apjfonts}
  \setcounter{secnumdepth}{5}
  \usepackage{graphicx}
  \submitted{27 November 2000}
}

\lefthead{HEGER ET AL.}
\righthead{}
\def\gtaprx {\lower .1ex\hbox{\rlap{\raise .6ex\hbox{\hskip .3ex
	{\ifmmode{\scriptscriptstyle >}\else
		{$\scriptscriptstyle >$}\fi}}}
	\kern -.4ex{\ifmmode{\scriptscriptstyle \sim}\else
		{$\scriptscriptstyle\sim$}\fi}}\xspace}
\def\ltaprx {\lower .1ex\hbox{\rlap{\raise .6ex\hbox{\hskip .3ex
	{\ifmmode{\scriptscriptstyle <}\else
		{$\scriptscriptstyle <$}\fi}}}
	\kern -.4ex{\ifmmode{\scriptscriptstyle \sim}\else
		{$\scriptscriptstyle\sim$}\fi}}\xspace}

\newcommand{\g}{{\ensuremath{\mathrm{g}}}\xspace}
\newcommand{\K}{{\ensuremath{\mathrm{K}}}\xspace}
\newcommand{\cm}{{\ensuremath{\mathrm{cm}}}\xspace}

\newcommand{\km}{{\ensuremath{\mathrm{km}}}\xspace}
\newcommand{\Msun}{{\ensuremath{\mathrm{M}_{\odot}}}\xspace}
\newcommand{\Sec}{{\ensuremath{\mathrm{s}}}\xspace}
\newcommand{\erg}{{\ensuremath{\mathrm{erg}}}\xspace}

\newcommand{\kms}{{\ensuremath{\km\,\Sec^{-1}}}\xspace}

\newcommand{\gcc}{{\ensuremath{\g\,\cm^{-3}}}\xspace}

\newcommand{\kB}{{\ensuremath{k_{\mathrm{B}}}}\xspace}

\newcommand{\Seci}{{\ensuremath{\Sec^{-1}}}\xspace}

\newcommand{\ergg}{{\ensuremath{\erg\,\g^{-1}}}\xspace}
\newcommand{\erggs}{{\ensuremath{\erg\,\g^{-1}\,\Sec^{-1}}}\xspace}

\newcommand{\Sunit}{{\ensuremath{\kB\,\mathrm{baryons}^{-1}}}\xspace}

\newcommand{\tb}{{\ensuremath{t_{\mathrm{b}}}}\xspace}
\newcommand{\tpresn}{{\ensuremath{t_0}}\xspace}

\newcommand{\Ye}{{\ensuremath{Y_{\mathrm{e}}}}\xspace}
\newcommand{\Se}{{\ensuremath{S_{\mathrm{e}}}}\xspace}
\newcommand{\St}{{\ensuremath{S_{\mathrm{tot}}}}\xspace}
\newcommand{\Yp}{{\ensuremath{Y(\mathrm{p})}}\xspace}

\newcommand{\snucq}{{\ensuremath{\epsilon_{\mathrm{nuc}}}}\xspace}
\newcommand{\snuw}{{\ensuremath{\epsilon_{\nu,\mathrm{weak}}}}\xspace}
\newcommand{\snubps}{{\ensuremath{\epsilon_{\nu,\mathrm{plas}}}}\xspace}
\newcommand{\sdotq}{{\ensuremath{\epsilon}}\xspace}

\newcommand{\sig}{{\ensuremath {S}}\xspace}
\newcommand{\sige}{{\ensuremath{S_{\mathrm{e}}}}\xspace}
\newcommand{\sigi}{{\ensuremath{S_{\mathrm{ion}}}}\xspace}
\newcommand{\sigr}{{\ensuremath{S_{\mathrm{rad}}}}\xspace}
\newcommand{\sigp}{{\ensuremath{S_{\mathrm{pair}}}}\xspace}

\newcommand{\lSect}[1]{{\label{sec:#1}}}
\newcommand{\lFig}[1]{{\label{fig:#1}}}

\newcommand{\lTab}[1]{{\label{tab:#1}}}
\newcommand{\pFig}[1]{{\placefigure{fig:#1}}}

\newcommand{\Tabff}[1]{{\ref{tab:#1}}}
\newcommand{\Tab}[1]{{Table~\Tabff{#1}}}
\newcommand{\Tabs}[1]{{Tables~\Tabff{#1}}}
\newcommand{\pan}[1]{{\textit{#1}}}

\newcommand{\FIGFF}[2]{{\ref{fig:#2}\pan{#1}}}
\newcommand{\Figff}[1]{{\FIGFF{}{#1}}}
\newcommand{\FIG}[2]{{Fig.~\FIGFF{#1}{#2}}}
\newcommand{\Fig}[1]{{\FIG{}{#1}}}
\newcommand{\FIGS}[2]{{Figs.~\FIGFF{#1}{#2}}}
\newcommand{\Figs}[1]{{\FIGS{}{#1}}}

\newcommand{\Sectff}[1]{{\ref{sec:#1}}}
\newcommand{\Sect}[1]{{Sect.~\Sectff{#1}}}

\newcommand{\I}[2]{{\isotope{}{#1}{#2}}}

\newcommand{\Ep}[1]{{\ensuremath{10^{#1}}}}
\newcommand{\E}[1]{{\ensuremath{\powersep\Ep{#1}}}}

\newcommand{\EC}{{\ensuremath{\mathrm{EC}}}\xspace}
\newcommand{\PC}{{\ensuremath{\mathrm{PC}}}\xspace}
\newcommand{\bd}{{\ensuremath{\beta^-}}\xspace}
\newcommand{\pd}{{\ensuremath{\beta^+}}\xspace}
\newcommand{\ed}{{\bd}\xspace}

\newcommand{\rate}{{\ensuremath{\lambda}}\xspace}

\newcommand{\rtotp}{{\ensuremath{\rate_{\mathrm{p,tot}}}}\xspace}
\newcommand{\wrate}{{\ensuremath{\rate_{\mathrm{weak}}}}\xspace}
\newcommand{\rectot}{{\ensuremath{\rate_{\EC,\mathrm{tot}}}}\xspace}
\newcommand{\rpdtot}{{\ensuremath{\rate_{\pd,\mathrm{tot}}}}\xspace}
\newcommand{\redtot}{{\ensuremath{\rate_{\bd,\mathrm{tot}}}}\xspace}

\newcommand{\Eweak}{{\ensuremath{E_{\mathrm{weak}}}}\xspace}
\newcommand{\Eplas}{{\ensuremath{E_{\mathrm{plas}}}}\xspace}
\newcommand{\EEC}{{\ensuremath{E_{\mathrm{\EC}}}}\xspace}
\newcommand{\Eed}{{\ensuremath{E_{\mathrm{\ed}}}}\xspace}

\newcommand{\F}{{\ensuremath{F}}\xspace}
\newcommand{\FEC}{{\ensuremath{\F_{\mathrm{\EC}}}}\xspace}
\newcommand{\Fpd}{{\ensuremath{\F_{\mathrm{\pd}}}}\xspace}
\newcommand{\Fed}{{\ensuremath{\F_{\mathrm{\ed}}}}\xspace}

\ifthenelse{\boolean{emul}}{\renewcommand{\pFig}[1]{}}

\newcommand{\FigOneFile}{fig1.ps} \newcommand{\FigOne}{Differences
  between the average neutrino energies emitted during electron
  capture for the new calculations of LMP and the old ones of FFN for
  several isotopic chains. The upper panel reflects the conditions
  during silicon core burning while the lower panel reflects the
  conditions during core contraction. In general, though not
  universally, the neutrino energies are larger in the new rate set.
  The entropy and energy losses associated with a certain amount of
  electron capture will thus be greater in the new models. \lFig{nue}}

\newcommand{\FigTwoAFile}{fig2a.ps}
\newcommand{\FigTwoBFile}{fig2b.ps} \newcommand{\FigTwo}{Histories of
the central temperature (\textit{thin lines}) and density
(\textit{thick lines}) in 15 and 25\,\Msun presupernova stars
calculated using the rate set of \cite{WW95} (WW; \textit{dotted
lines}), the full implementation of FFN rates (electron capture and
beta decay; \textit{solid lines}) and the new weak interaction rates
of Langanke \& Mart\'{\i}nez-Pinedo (LMP; \textit{dashed lines}).
\lFig{trho}}

\newcommand{\FigThreeFile}{fig3.ps} 
\newcommand{\FigThree}{Evolution of \Ye with time at the center of
five 15\,\Msun stars that used a) the new rates (LMP; dark solid
line); b) the full implementation of FFN rates including beta-decay
(light solid line; and c) the weak rates used by WW (dot-dashed line).
Also shown for comparison are the results using the LMP rates with
beta-decay rates multiplied by zero and two.  The WW models are very
similar to what is obtained when beta-decay is neglected.  The final
central values of \Ye for the models are in case a) 0.432; b) 0.430;
and c) 0.423 (\Tab{15}).
\lFig{yeoft15}}

\newcommand{\FigFourFile}{fig4.ps} \newcommand{\FigFour}{Same as
\Fig{yeoft15}, but for 25\,\Msun models. Here the revised rates for
electron capture and the inclusion of beta decay have comparable
effects on the central evolution.  The final central values of \Ye
for the models are in case a) 0.445 for LMP rates; b) 0.438 for FFN
rates; and c) 0.430 for the WW model.  
\lFig{yeoft25}}

\newcommand{\FigFiveFile}{fig5.ps} 
\newcommand{\FigFive}{Same as \Fig{yeoft15}, but for 40\,\Msun
models. As for the 25\,\Msun star (\Fig{yeoft25}), the revised rates
for electron capture and the inclusion of beta decay have comparable
effects on the central evolution.  The final central values of \Ye
for the models are a) 0.447 for LMP rates, b) 0.439 for FFN rates, 
and c) 0.430 for the WW models.
\lFig{yeoft40}}

\newcommand{\FigSixFile}{fig6.ps} 
\newcommand{\FigSix}{Partial contributions to the evolution of the
central electron mole number, \Ye, as a function of time until
dynamic collapse in a 15\,\Msun star.  The calculation on the left
used the new LMP rates.  A calculation using the full
implementation of the FFN rates would look very similar. Note the
initial dominance of electron capture, but a period around $\log
(\tb - t) = 3.5$ to $4.5$ when beta-decay balances electron
capture. Still later, the increasing density in the contracting core
favors electron capture again and beta-decay cannot keep up. However,
time has become so short that \Ye changes very little in these last
few hours (\Fig{yeoft15}).  The right hand figure shows the calculation
of WW where beta equilibrium was not achieved.  The shaded region is 
the epoch of convective silicon core burning.
\lFig{lmp15}}

\newcommand{\FigSevenFile}{fig7.ps}
\newcommand{\FigSeven}{Same as \Fig{lmp15}, but for 25\,\Msun models
and the comparison is with respect to the full FFN implementation
rather than WW.  Using the new LMP rates, beta-decay competes with
electron-capture, but never quite balances it. Using FFN rates, beta
equilibrium is achieved. A plot for the 40\,\Msun models is very
similar.  A plot of the calculation using the WW rates resembles
\Fig{lmp15}.  The shaded region is 
the epoch of convective silicon core burning.
\lFig{lmp25}}

\newcommand{\FigEightFile}{fig8.ps} 
\newcommand{\FigEight}{Time-weighted weak flows in the 25\,\Msun
model using the weak rates of LMP. Most of the change in \Ye is
consequence of electron capture though beta decay becomes important at
late times.  \lFig{linearlmp}}

\newcommand{\FigNineFile}{fig9.ps} \newcommand{\FigNine}{Major
contributions to the electron capture flow for two 15\,\Msun stars
that used the LMP rates (left) and the FFN rates (right). For the FFN
rates, the evolution after $\log(\tpresn-t) = 5$ is dominated by
electron capture on $^{60}$Co. Electron capture is unimportant. For
the LMP rates, electron capture on $^{60}$Co is unimportant, but
capture on free protons is beginning to contribute in the presupernova
model, a trend that will probably continue as the star
collapses. During the pre-collapse evolution though, most of the flow
in the LMP models is carried by electron-capture on other heavy
nuclei, especially $^{54,55,56,57}$Fe, $^{55}$Mn, and $^{61}$Ni
(\Tab{15}).  The shaded region is the epoch of convective silicon core
burning.  \lFig{contrib15}}

\newcommand{\FigTenFile}{fig10.ps} \newcommand{\FigTen}{Major
contributions to the electron capture flow for two 25\,\Msun stars
that used the LMP rates (left) and the FFN rates (right). See also
\Fig{contrib15}.  Again the evolution of \Ye in the model using the
FFN rates is almost completely determined by
$^{60}$Co(e$^-,\nu)^{60}$Fe.  The evolution of the model with LMP
rates on the other hand is determined by other nuclei, especially
$^{53,54,55,56}$Fe, $^{55}$Co and $^{56}$Ni (\Tab{25}) with an
increasingly important contribution from electron capture on free
protons.  A graph for the 40\,\Msun model looks similar, except that
at collapse p(e$^-,\nu)$n contributes almost the entire flow.  The
shaded region is the epoch of convective silicon core burning.
\lFig{contrib25}}

\newcommand{\FigElevenFile}{fig11.ps} 
\newcommand{\FigEleven}{Structure of the 15\,\Msun presupernova star as
calculated by WW and recalculated using the new LMP rates. Note the
lower central entropy in the LMP model.  The discontinuities in $T$, \Ye,
and entropy at 0.98\,\Msun in the LMP model and 0.92\,\Msun in the
WW model \ show the extent of the first stage of silicon core
burning. In the FFN model (not shown) this discontinuity is located at
0.98\,\Msun.  The abrupt increase in \Ye at 1.29\,\Msun in all three
models marks the maximum extent of the silicon convective shell and
the edge of the iron core.  The total entropy and its components are
shown.  The LMP rates give a lower entropy both in the center
and outer core.  The entropy jump at 1.43\,\Msun in all the models is
the location of the oxygen burning shell (edge of the silicon core).
\lFig{15presn}}

\newcommand{\FigTwelveFile}{fig12.ps}
\newcommand{\FigTwelve}{Structure of the 25\,\Msun presupernova.
Similar to \Fig{15presn}, but the decrease in central entropy for the
LMP based model is less pronounced.  The central entropy for the FFN
model (not shown) is smaller than either of these.  The maximum extent
of the silicon convective core gives fossil discontinuities in $T$,
\Ye, and the entropy at 1.07\,\Msun in the LMP model, 1.12\,\Msun in
the WW models, and 1.18\,\Msun in the FFN model.  The jump in \Ye at
1.72\,\Msun in the LMP model and 1.78\,\Msun in both the WW and FFN
models marks the maximum extent of the silicon convective shell and
the edge of the iron core.  The jump in entropy at 2.06\,\Msun in all
three models is the base of the oxygen shell (i.e., the edge of the
silicon core).  \lFig{25presn}}

\newcommand{\FigThirteenFile}{fig13.ps} \newcommand{\FigThirteen}{Central
vales of \Ye (\textbf{a}) and masses of the ``deleptonized core''
(\textbf{b}) at the time of iron core implosion for a finely spaced
grid of stellar masses.  The ``deleptonized core'' mass is defined as
the mass interior to the point where the \Ye is 0.49.  Despite the
larger values of \Ye for the new rates (Panel a), the deleptonized
core masses are not changed very much for the most common variety of
supernovae, those with masses below 20\,\Msun.  However larger changes
result for stars of higher mass.  \lFig{yefeofm}}

\newcommand{\FigFourteenFile}{fig14.ps}
\newcommand{\FigFourteen}{Total entropy at the center of a collection
of presupernova stars of varying mass.  \textsl{Crosses} indicate
models calculated with the new weak rates (LMP); \textit{circles} use
the rate set of WW.  For stellar masses below about 20\,\Msun the
entropy is systematically lower in the new models.  From 20 to
27\,\Msun the entropy is not changed very much, though there are
exceptions at 25 and 26\,\Msun.  Above 27\,\Msun, the trend seen at
lower masses reverses itself and the new models have higher total
\emph{central} entropy. However to see the effect of entropy on the
iron core mass, one must really consider the distribution of
electronic entropy in the entire iron core (\Figs{15presn} and
\Figff{25presn}).  \lFig{entropy}}

\newcommand{\FigFifteenFile}{fig15.ps}
\newcommand{\FigFifteen}{Evolution of the central entropy in 
25\,\Msun models using weak rates from FFN, LMP, and WW.  The bump
starting at $\log(\tpresn-t) = 4.7$ is due to core silicon burning
(\Fig{trho}).  In general, the advanced stages of nuclear burning in
the cores of massive stars cause rises in entropy followed by declines
as neutrinos cool the depleted region while shell burning supports the
star.  The model of WW tracks the FFN curve until beta decay becomes
important, then coincidentally follows the LMP trajectory. At late
times, the entropy using the full FFN rate set is lower because
continued beta equilibrium leads to additional cooling (\Fig{lmp25}).
\lFig{entropyoft}}

\newcommand{\FigSixteenFile}{fig16.ps} 
\newcommand{\FigSixteen}{Silicon cores for the two sets of weak rates.
The silicon core is defined as the point where the oxygen abundance declines to
very small values.  Typically there is a large entropy jump associated with the 
base of the oxygen shell so that the silicon core may be as relevant for
the explosion mechanism as the iron core mass.  The weak rates do not 
have a great effect on the mass of the silicon core.  
\lFig{sicore}}

\slugcomment{Submitted ApJ, 27 November 2000} 

\begin{document}

\title{Presupernova Evolution with Improved Rates for Weak Interactions}
\author{A. Heger, S. E. Woosley, G. Mart\'{\i}nez-Pinedo, and K. Langanke}

\vskip 0.2 in
\affil{Department of Astronomy and Astrophysics \\
University of California, Santa Cruz, CA 95064\\[2mm]
 Institut for Fysik og Astronomi\\
 \AA{}rhus Universitet, DK-8000 \AA{}rhus C, Denmark}
\authoremail{alex@ucolick.org}

\begin{abstract} 
Recent shell-model calculations of weak-interaction rates for nuclei
in the mass range $A = 45-65$ have resulted in substantial revisions
to the hitherto standard set of Fuller, Fowler, \& Newman (FFN).  In
particular, key electron-capture rates, such as that for \I{60}{Co}
are much smaller.  We consider here the effects of these revised rates
on the presupernova (post-oxygen burning) evolution of massive stars
in the mass range 11 to 40\,\Msun.  Moreover, we include, for the first
time in models by our group, the effects of modern rates for
beta-decay in addition to electron capture and positron emission.
Stars of 15, 25, and 40\,\Msun stars are examined in detail using
both the full FFN rate set and the revised rates.  An additional
finely spaced (in mass) grid of 34 models is also calculated in order
to give the systematics of iron core and silicon core masses. Values
for the central electron mole number at the time of iron core collapse
in the new models are typically larger, by $\Delta \Ye = 0.005$ to
0.015, than those of \cite{WW95}, with a tendency for the more massive
models to display larger differences. About half of this change is a
consequence of including beta-decay; the other half, result of the
smaller rates for electron capture. Unlike what might be expected
solely on basis of the larger \Ye's, the new iron core masses are
systematically smaller owing to a decrease in the entropy in the outer
iron core.  The changes in iron core mass range from zero to
0.1\,\Msun (larger changes for high mass stars). It would be
erroneous though to estimate the facility of exploding these new
models based solely upon their iron core mass since the entire core
structure is altered and the density change is not so different as the
adjustments in composition might suggest. We also observe, as
predicted by \cite{auf94a}, a tendency towards beta-equilibrium just
prior to the collapse of the core, and the subsequent loss of that
equilibrium as core collapse proceeds.  This tendency is more
pronounced in the 15\,\Msun model than in the heavier stars. We
discuss the key weak reaction rates, both beta-decay and
electron-capture, responsible for the evolution of \Ye and make
suggestions for future measurements.
\end{abstract}



\section{Introduction}
\lSect{intro}

The effect of weak interactions on the final evolution of massive
stars is known to be quite pronounced \cite[e.g.,][]{WFW84}.  Weak
interactions alter the composition and change the nucleosynthesis.
They modify the structure of the iron core and its overlying mantle
and thus help to determine whether the star explodes and the
characteristics of the bound remnant. Electron capture decreases the
number of electrons available for pressure support and, all else being
equal, decreases the effective Chandrasekhar Mass that can be
supported without collapsing. Beta decay acts in the opposite
direction. Weak interactions also create neutrinos that, for all
densities until the collapse becomes truly hydrodynamic (i.e., $\rho
\ltaprx 10^{11}$ g cm$^{-3}$), escape the star carrying away energy
and thereby reducing entropy.  In fact, for stars that are not too
massive, neutrinos from weak interactions, as opposed to thermal
plasma processes, are the dominant energy sink in the presupernova
star \cite[]{WZW78}.  Reducing the entropy also makes the iron core
smaller \cite[]{Timmes96}.

Over the years, many people have calculated weak interaction rates for
astrophysical application
\cite[]{Hans66,Hans68,Mazu73,MTC74,Taka73,auf94b}.  For approximately
15 years though, the standard in the field has been the tabulations of
\cite{FFN80,FFN82a,FFN82b,FFN85} (henceforth FFN).  These
authors calculated rates for electron capture, beta-decay, and
positron emission, plus the associated neutrino losses for all
astrophysically relevant nuclei ranging in atomic mass from 23 to 60.
Their calculations were based upon an examination of all available
information (in the mid 1980's) for individual ground states and
low-lying excited states in the nuclei of interest.  Recognizing that
this only saturated a small part of the Gamow-Teller resonance, they
extrapolated the known experimental rates using a simple
(single-state) representation of this resonance.

Now, as we briefly summarize in \Sect{newrates}, new shell models of
the distribution of Gamow-Teller strength have resulted in an improved
- and often reduced - estimate of its strength.  The alteration in
some important rates, such as electron capture on $^{60}$Co, is as
large as two orders of magnitude in important astrophysical
circumstances.  In order to determine how these new rates affect the
post-oxygen burning evolution of massive stars and presupernova
models, we have examined the evolution of a fine grid of 34 stellar
models spanning a mass range 11 to 40\,\Msun.  The post-oxygen burning
evolution of each star was calculated twice - once using the same old
set of weak rates as employed in \cite{WW95} (henceforth WW) by Tom
Weaver in 1996 (priv. com.; unpublished) and, again, with the new
rates.  It is important to note that the models of WW included
electron capture and positron decay rates from FFN, but used
beta-decay rates from \citeauthor{Hans68} and \citeauthor{Mazu73}.
These were much smaller than their FFN counterparts and essentially
resulted in beta decay being neglected.  This was a mistake, that, so
far as we are aware, was not replicated in the more recent works of
other groups \cite[]{Thie96,Chie98}.  An examination of conditions in
the pre-supernova model led to the (correct) conclusion that
beta-decay was negligible there. What was overlooked however, as was
subsequently pointed out \cite[]{auf94a}, was the importance of
beta-decay in the iron core a few hours \emph{before} iron core
collapse.

The new weak rate implementation described here \emph{does} include
beta-decay - along with revised rates for electron capture and
positron emission. In order to make a fair comparison to what would
have been obtained by WW had they used the full FFN implementation, we
have recalculated (\Sect{stars}) three stars (15, 25, and 40\,\Msun)
using FFN beta decay rates as well as FFN electron-capture and
positron-decay rates keeping the stellar physics (thermal neutrino
loss rates, opacities, mass loss, etc.) the same as in WW.  Subsequent
survey of massive stellar evolution \cite[]{Raus00,Hege01} will
implement the full revised weak rate set along with many other
improvements in a fine grid of masses and metallicities.

We find (\Sect{newresults}) significant differences in the
presupernova models calculated using the new rates In particular, the
large decrease in key rates for electron-capture and the inclusion of
beta-decay result in an electron mole number, \Ye, in the center of
the presupernova star that is 1\,\% to 4\,\% larger. Changes around 3\,\%
are most common and there is a tendency for the biggest changes to
occur for larger stellar masses.  The altered weak rates also affect
the structure of the presupernova star, leading to iron core masses
that are, almost without exception, \emph{smaller}.  This seemingly
paradoxical result (smaller Fe core and larger \Ye) comes about
because of a decreased entropy in the outer iron core, a consequence
of a final stage of silicon shell burning.

In \Sect{conclude} we summarize our findings and enumerate key rates
that warrant further laboratory study (see also \Tabs{15} --
\Tabff{flow40}).

\section{The New Rates}
\lSect{newrates}

Weak interactions in presupernova stars are known to be dominated by
allowed Fermi and Gamow-Teller transitions \cite[]{bet79}.  FFN
estimated their rates based upon the presence of a single Gamow-Teller
(GT) resonance which they derived on the basis of the independent
particle model, supplemented, where available, by experimental data
for low-lying transitions.  However, recent experimental data on the
GT distributions in nuclei \cite[]{alf90,vet89,alf93,elk94,wil95}
shows a quenching of the total GT strength compared to the
independent-particle model value. More importantly, the data also show
a fragmentation of the GT strength over many states in the daughter
nucleus. Due to the strong dependence of the phase space on the
energy, particularly for stellar electron capture, an improved
treatment of the relevant nuclear structure is called for
\cite[]{auf93a,auf93b}.

The origin of the quenching and fragmentation of the GT strength
distribution has been traced to nucleon-nucleon correlations, in
particular, correlations between protons and neutrons.  Furthermore,
studies of $sd$-shell nuclei ($A=17-39$) by \cite{BW88} have shown
that reproduction of the observed GT strength distributions requires
the inclusion of all nucleon-nucleon correlations within the complete
shell.  This requires the shell model. Following the work by Brown and
Wildenthal, \cite{oda94} calculated shell-model rates for all the
relevant weak processes for $sd$-shell nuclei. Complete shell model
calculations for the heavier $pf$-shell nuclei were deferred because
these require model spaces orders of magnitude larger.  Such
calculations only became possible after significant progress both in
shell-model programming and computer power.

Today, shell-model calculations for the $pf$-shell can be performed at
a truncation level which guarantees convergence. \cite{cau99}, in
particular, have demonstrated that state-of-the-art diagonalization
studies, typically involving a few times 10 million configurations,
can reproduce all the relevant ingredients - the GT$_\pm$ strength
distributions for changing protons (neutrons) into neutrons (protons),
the level spectra, and the half lives - and hence have the predictive
power to calculate accurate stellar weak interaction rates for nuclei
up to A = 65. A survey of the $pf$-shell nuclei has been completed by
\cite{LM00} and is ready for incorporation in stellar models. In this
paper we combine the rate sets of Oda et al. (up to A = 40) with those
of Langanke and Mart\'{\i}nez-Pinedo for A = 45 to 65. Shell-model rates
for nuclei with $A=40-44$ are not yet available, as these nuclei
require the inclusion of both $sd$- and $pf$-shells. In our studies we
have used the FFN rates for these nuclei. We shall refer to the hybrid
collection as the ``LMP rate set''.  Though it clearly involves
multiple sources, the most important, uncertain rates for the late
stages of stellar evolution involve the $pf$-nuclei.  Though these new
shell-model rates and the FFN rates agree rather well for $sd$-shell
nuclei \cite[]{oda94}, there are appreciable differences at higher
mass.  The electron-capture rates, and to a lesser extent, the
$\beta$-decay rates \cite[see][]{MLD00} are smaller than the FFN
rates by, on average, one order of magnitude.  The causes have been
discussed by \cite{LM00}.

We include in our studies rates for electron and positron captures and
$\beta^+$ and $\beta^-$ decays. Energy losses by neutrinos, and hence
core cooling, are determined from the (tabulated) average neutrino
energies which have been consistently calculated from the shell-model
studies for each of the weak processes.  In addition to the rates
themselves, the energy losses to neutrinos can have an important
effect on the stellar structure, particularly in the post-silicon
burning phase (neutrino losses {\sl during} silicon burning are
dominated by thermally produced neutrinos).  \Fig{nue} shows the mean
neutrino energies emitted during electron capture on various parent
nuclei using the LMP and FFN tabulations. There is considerable
variation in the ratio, but a significant tendency for the neutrinos
emitted by weak interactions in the post-silicon burning phase (e.g.,
$T_9 = 5$) to have higher mean energies in the LMP rate set.

\ifthenelse{\boolean{emul}}{
\vspace{1.5\baselineskip}
\noindent
\includegraphics[width=\columnwidth]{\FigOneFile}
\figcaption{\FigOne}
\vspace{1.5\baselineskip}}{}

\section{Stellar Models}
\lSect{stars}

All models were calculated using a modified version of the KEPLER
implicit hydrodynamics code \cite[]{WZW78}.  In order to facilitate
comparisons with earlier work, recent improvements to the code -- the
inclusion of OPAL opacities, mass loss, rotationally induced mixing,
and recent revisions to nuclear reaction rates (other than weak rates)
were not included.  A total of 34 models were calculated.  The
procedure and models were as close as we could manage to WW.
Pre-explosive hydrogen, helium, carbon, and neon burning and the early
stages of central oxygen burning should not be affected by the current
revision of the weak rates and so all studies commenced from
previously existing ``binary restart dumps'' that were made when the
central oxygen abundance was depleted, by oxygen burning, to a mass
fraction $X(\I{16}O)=0.04$.  The evolution was then followed through
the remainder of oxygen burning, silicon core and shell burning, and
the early stages of collapse.  The ``presupernova model'' was defined
as the time when the collapse velocity near the edge of the iron core
first reached 1000 km s$^{-1}$, as in \cite{WW95}.  In the remainder
of the paper we refer to this time as \tpresn.

\ifthenelse{\boolean{emul}}{
\vspace{1.5\baselineskip}
\noindent
\includegraphics[width=\columnwidth]{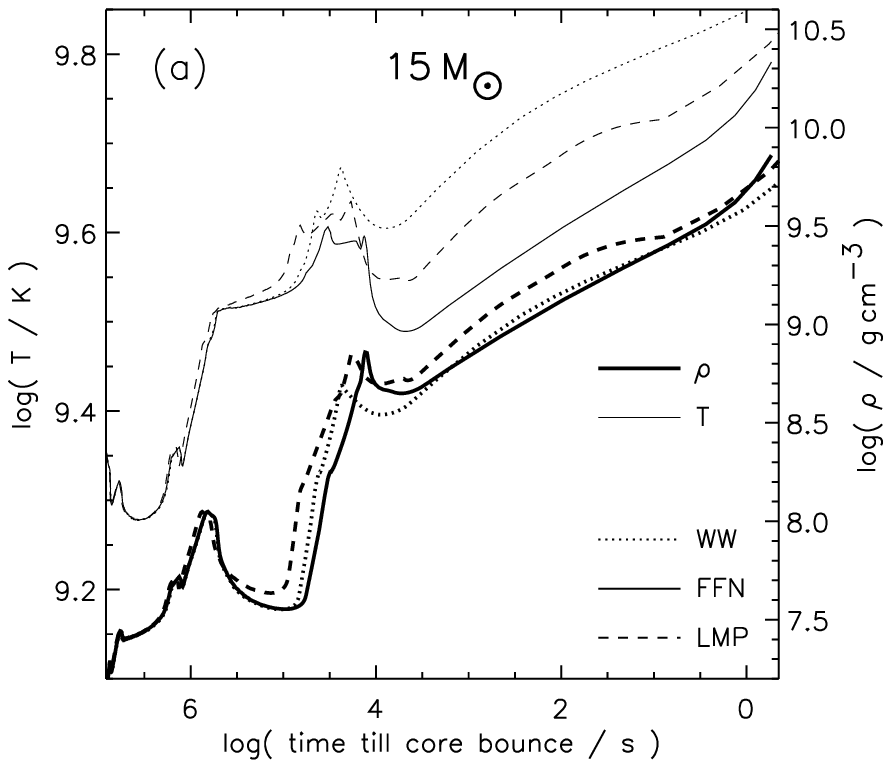}
\hspace{0.24 in}
\includegraphics[width=\columnwidth]{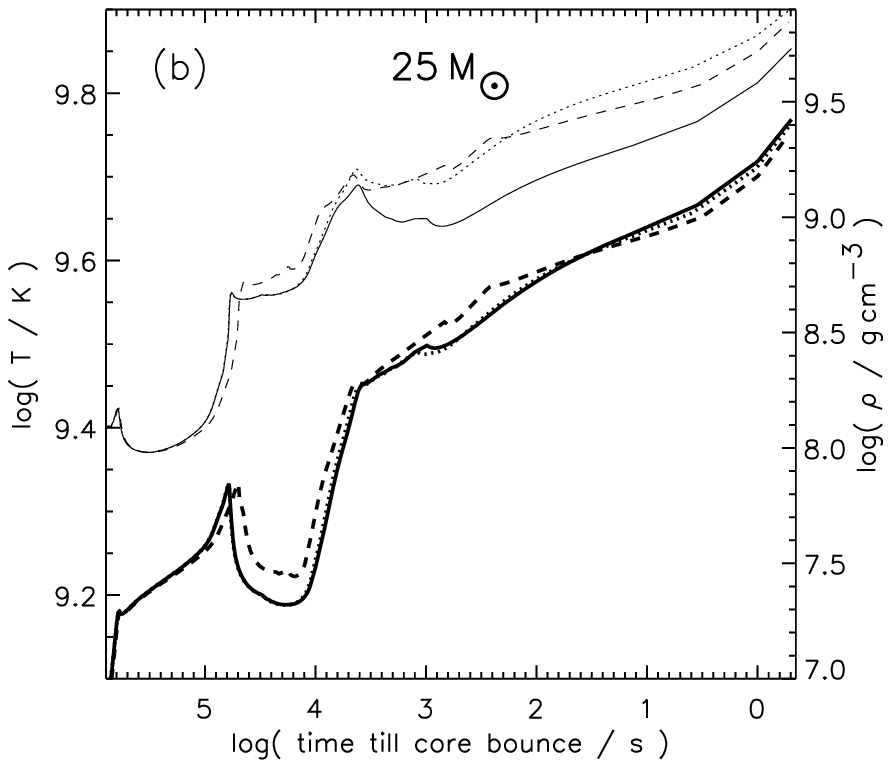}
\figcaption{\FigTwo}
\vspace{1.5\baselineskip}}{}

Three versions of KEPLER were employed. One was the same as in WW and
was used to duplicate the earlier calculations in order to obtain
additional edit quantities for comparison.  The second used the new
rate set described in \Sect{newrates}.  These included rates for
electron capture, beta decay, and positron emission from \cite[]{LM00}
(henceforth LMP) for nuclei heavier than mass 45 and from \cite{oda94}
and FFN for nuclei lighter than mass 45, as explained above.  We shall
find that it is the nuclei above mass 45 that determine \Ye and the
presupernova structure.  The effects of positron capture were
estimated to be small for the stars under consideration here (though
not perhaps for much more massive stars) and were \emph{not} included
in our calculations.  In order to preserve accuracy using a table with
a limited number of temperature and density points, an effective $ft$
value was determined for the electron capture rates by dividing out
the first order dependence on Fermi integrals (this was also done with
the FFN rates in prior work by WW).  The rates for beta decay and
positron emission, however, were interpolated logarithmically (using a
cubic spline) in the tables provided.  The mean energy of neutrinos
emitted in each reaction are also given by LMP, \citeauthor{oda94},
and FFN.  These were also interpolated logarithmically. A third
version of KEPLER was the same as WW except for the incorporation of
FFN beta decay. Where results are given in this paper labelled
``FFN'', they refer to calculations using this third version of
KEPLER.  Similarly, ``LMP'' refers to the second and ``WW'' to the
first.

All three versions were used to study 15, 25, and 40\,\Msun
presupernova models and these are the three stars that will be used
for illustration in this paper.  The presupernova models from these
calculations are available to those wishing to study their further
(explosive) evolution. An additional set of 31 other masses were also
studied in the 11 - 40\,\Msun range. Details of these models,
calculated by Tom Weaver in 1995 and 1996, remain unpublished though
they were discussed and used in \cite{Timmes96} and
employed identical physics to those in WW.  Oxygen depletion dumps were
available for these stars and a second version of the presupernova
evolution of each was calculated using the LMP rates.  The comparison
of results for these additional models will be useful later in the
paper where systematics of core entropy, iron core mass, silicon core
mass, and \Ye are discussed. 

\Fig{trho} shows the evolution of the 15 and 25\,\Msun models using
the WW rates (WW), the FFN rates (FFN), and the new weak rates (LMP)
and defines the temperature-density region of interest in this study.
The time axis in this and subsequent figures warrants some
explanation. WW defined ``presupernova'', somewhat arbitrarily, as
that point in the final evolution of a massive star when any point in
its iron core reached a collapse velocity in excess of
1000\,\kms. Operationally, this was a good time to link the results of
a stellar evolution model to a calculation of the explosion mechanism
since the codes used to study the explosion could usually only follow
dynamic evolution. Another fiducial time is that instant, which we
shall call the ``bounce time'', \tb, when the iron core collapses to
supra-nuclear density and rebounds creating a shock wave.  This might
be more properly regarded as the end of the star's life and the
beginning of the supernova.  We estimated \tb by running the KEPLER
calculation until the central density exceeded \Ep{12}\,\gcc.  This is
within a millisecond or so of the core bounce.  We used this estimate
of \tb in all our graphs and tables.  Because of the accelerating
evolution of the star, evolving quantities are best plotted against
the logarithm of the time remaining until \tb.

\section{Results and Discussion}
\lSect{newresults}

\Tabs{15} -- \Tabff{40} summarize some interesting physical quantities
at representative times in the fiducial set of 15, 25, and 40\,\Msun
stars.  Though three sets of models (WW, FFN, and LMP) were studied,
only two are included in the tables.  As we shall see, beta decay
remains unimportant until where after the depletion of silicon in the
center of the star.  Thus the results labeled FFN are equivalent to WW
until the star reaches silicon shell burning. Afterwards, it is more
meaningful to compare results using the new rate LMP set to what would
have been obtained using an equivalent full implementation of the FFN
rates than comparing to the calculations of WW.

The times chosen for edit in \Tabs{15} -- \Tabff{40} are somewhat
arbitrary but attempt to capture the essence of the evolutionary state
of the center of the star at equal times before core bounce and at
times representative of oxygen depletion (``O-dep''); oxygen shell
burning ``O-shell''; the ignition of silicon in the core (first
convection; ``Si-ign''); half-way through silicon burning
(``Si-burn''); the end of convective core silicon burning
(``Si-dep''); the ignition of the first stage of convective silicon
shell burning (``Si-shell''); after silicon shell extinction (``core
contr''); and the presupernova star as defined in \Sect{stars}.  In
\Tabs{15} -- \Tabff{40}, the relation between ``presupernova'' time
and \tb, the core bounce time, is clearly stated.

In addition to the values of the central temperature, density, and
\Ye, key energy generation and loss terms, entropies, and weak flows
are also edited.  The specific central nuclear energy generation
rates, the weak and plasma neutrino losses and the total energy
``generation'' rate is given next.  Then we give the total central
entropy and its contributions by electrons, ions, radiation and pairs,
and in the next two lines the proton mole fraction and the rate of
electron capture onto protons.  The last row before the horizontal
line gives the total net weak rate.  Below that line we give first the
five most important ions that \emph{decrease} \Ye (electron capture
and \pd-decay) and their rates (below the ion symbol) and then sum up
the total rates of electron capture and \pd-decay.  In the last part
of the table we first give the five most important nuclei that
\emph{increase} \Ye (positron capture and \bd-decay) and in the last
line the total rate of \bd-decay. 

For comparison to the ``pre-SN'' values given in \Tabs{15} --
\Tabff{40} for the FFN and LMP rate sets, we note that the central
entropies at the equivalent times using the WW models were 0.73, 0.97,
and 1.10 for the 15, 25, and 40\,\Msun models respectively.

From these tables, one can infer many of the alterations that come
about as a consequence of the new rate set. However, there are a few
we want to emphasize.

\subsection{The evolution of \protect\Ye and the approach to beta equilibrium}
\lSect{betaequil}

\Fig{yeoft15}, \Fig{yeoft25}, and \Fig{yeoft40} show the evolution of
the central value of \Ye for different choices of rate set. For the
15\,\Msun model, which as we shall see briefly achieves full dynamic
beta-equilibrium, the choice of rate set is not so important as the
inclusion of beta-decay.  The old results of WW are essentially a lower
bound to the \Ye one expects when beta decay is zero. For the higher
mass stars, the actual size of the electron capture rates has a more
significant effect amounting to about half of the difference between
the WW models and the new models.

Most of the change in central \Ye in the 15\,\Msun models occurs
during the period \Ep6 to \Ep4 s before core collapse.  This is a
time (\Tab{15}) when the star is either burning silicon in its core
($\log (\tb -t) = 5.8$ to 4.8) or is resting in hydrostatic
equilibrium with a central core of iron while silicon burns in a shell
at about 1.1\,\Msun.  Capture rates on iron group nuclei are thus most
important. It is also interesting that the final \Ye is set to
within a percent or so well \emph{before} the iron core begins its final
Kelvin-Helmholtz contraction (the last hour).  This means that the most
important period for determining core structure, or at least the
electron fraction in the center, occurs, not during the dynamic
implosion of the star, but during silicon shell burning.

\Tab{15} also shows the importance of neutrino losses from weak
interactions during these final stages. For the 15\,\Msun model,
neutrino losses from electron capture and beta-decay exceed thermal
neutrino losses {\sl in the center of the star} during and after
silicon shell burning (but not in the shell itself). By the time of
the presupernova model, neutrino losses from weak flows are over four
orders of magnitude greater in the center of the star than those from
thermal neutrinos.  They are also competing with photodisintegration
in causing the collapse of the core.

For the 25 and 40\,\Msun models, the situation is similar though
thermal neutrinos are more important at later times and constitute
about 1\,\% of the central energy loss rate in the presupernova star.

\ifthenelse{\boolean{emul}}{
\vspace{1.5\baselineskip}
\noindent
\includegraphics[width=\columnwidth]{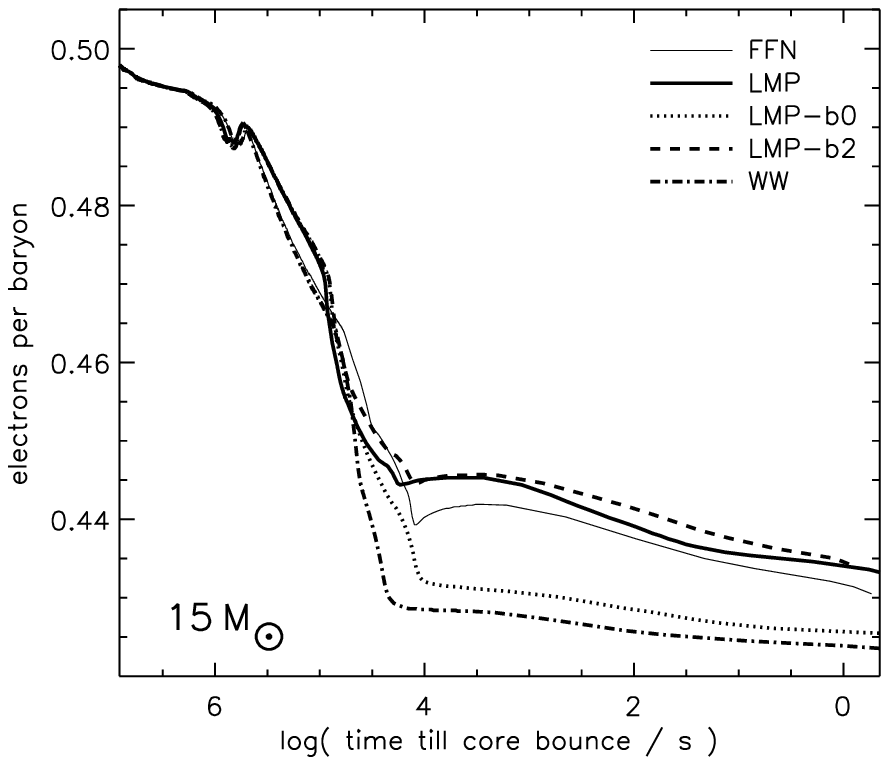}
\figcaption{\FigThree}
\vspace{1.5\baselineskip}}{}

\ifthenelse{\boolean{emul}}{
\vspace{1.5\baselineskip}
\noindent
\includegraphics[width=\columnwidth]{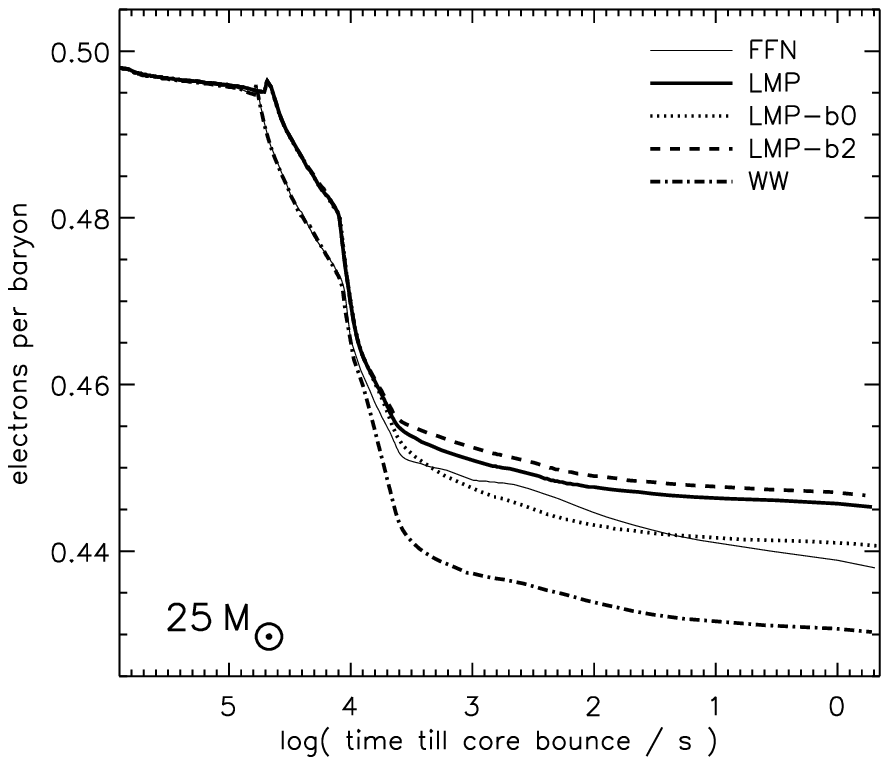}
\figcaption{\FigFour}
\vspace{1.5\baselineskip}}{}

\ifthenelse{\boolean{emul}}{
\vspace{1.5\baselineskip}
\noindent
\includegraphics[width=\columnwidth]{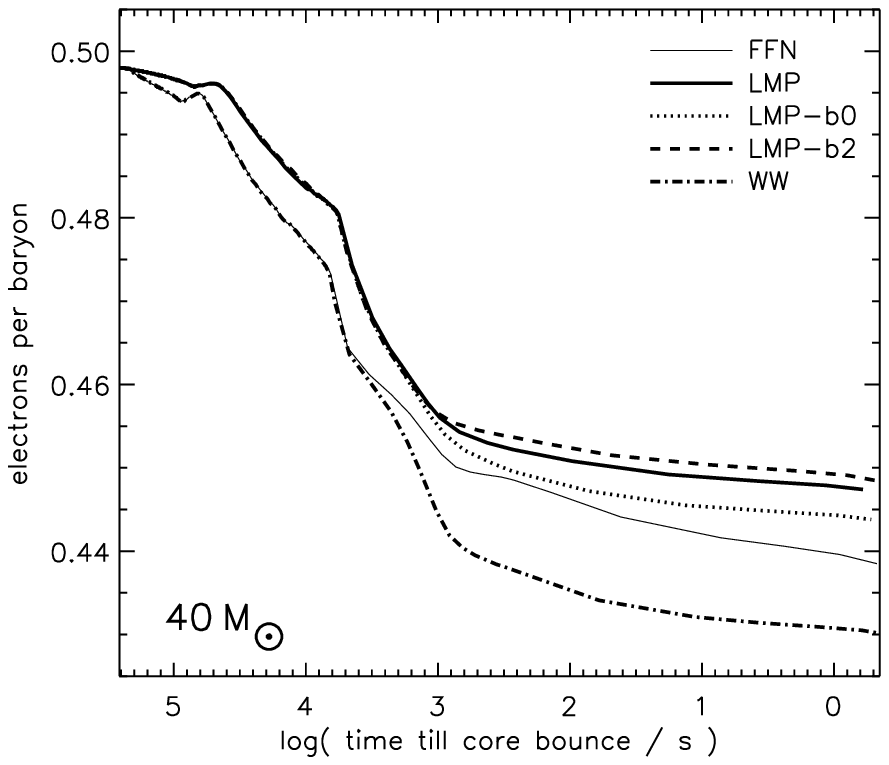}
\figcaption{\FigFive}
\vspace{1.5\baselineskip}}{}

An interesting phenomenon observed in all models that included modern
rates for beta-decay was the approach, during the late stages of
silicon shell burning and early iron core contraction, to dynamic
beta-equilibrium (\Tabs{15} -- \Tab{40}; \Fig{lmp15}, and
\Fig{lmp25}). Such a close approach was not observed in models using
the older rates (\Fig{lmp15}).  About 1 to 10 hours prior to core
collapse in the 15\,\Msun model, for example, the total rate of
beta-decay very nearly balances the total rate of electron
capture. Such a balance, as foreseen by \cite{auf94a}, is never quite
achieved, with the new rates, in the 25 and 40\,\Msun models, though
it comes closer with the rates of FFN. It is important to note that
this state of \emph{dynamic} equilibrium is not the same as true weak
equilibrium (such as exists in the early instants of the Big Bang). A
state of detailed balance does \emph{not} exist between all URCA
pairs.  The gas is still transparent to neutrinos and the reciprocal
processes of electron capture and beta decay, i.e., neutrino capture
and anti-neutrino capture, are quite negligible.  This makes it
impossible to assign a local chemical potential to the
neutrinos. Moreover, electron capture on a given nucleus, $^{\mathrm
A}$Z, including all states, ground and excited, does not occur at a
rate very nearly equal to the beta decay of $^{\mathrm
A}$(Z-1). Rather it is the integral of all beta decays that balances a
similar integral of all electron captures to render d\Ye/d$t$ close to
zero and \Ye stationary. Of course, evolution does occur as the
temperature and density evolve, and eventually this dynamic
beta-equilibrium ceases to exist.  ``Freeze-out'' happens as the
density rises to the point that electron capture is favored, but the
rates become too slow to maintain steady state.

Another feature of this dynamic equilibrium that distinguishes it from
true equilibrium is that the rates still matter.  The abundances are
not, in general, given simply by the Saha equation. A systematic
decrease in electron capture rates, for example, shifts the steady
state \Ye at a given temperature and density to a larger
value. However, the change is not great.  A larger \Ye implies a
greater abundance of nuclei that are less neutron-rich and whose
beta-decay rates will be smaller because of the lower $Q$-values.  The
distribution responds by decreasing \Ye to the point where,
globally, electron capture again balances beta decay. A similar
``restoring force'' occurs if \Ye is displaced to smaller
values. Since the $Q$-values enter as high powers in the phase space
integrals, there is, overall, a \emph{nearly} unique solution for \Ye
at a given temperature and density.  This explains, in part, the robust
value of the final \Ye in the 15\,\Msun model.  Though the important
electron capture rates are, at a given temperature, density, and
\Ye, quite a bit smaller in the new models, a small adjustment of
\Ye in steady state compensates for this so that the final value of
\Ye doesn't change much.  In the 25 and 40\,\Msun models where
dynamic beta equilibrium is not achieved, \Ye remains more sensitive
to the rates.

Despite the fact that electron capture flows are not, in general,
balanced by beta decay for each pair of nuclei, there are some
interesting URCA pairs apparent in \Tabs{15} -- \Tabff{40}.  For
example, during silicon shell burning and core contraction in the
15\,\Msun star (\Tab{15}), $^{57}$Fe(e$^-,\nu)^{57}$Mn and
$^{56}$Fe(e$^-,\nu)^{56}$Mn are two of the three most important
electron capture flows, but $^{57}$Mn(e$^- \bar \nu)^{57}$Fe and
$^{56}$Mn(e$^- \bar \nu)^{56}$Fe are also two of the three most
important beta-decay flows.  The rates for these forward and inverse
reactions are calculated independently, but perhaps there is some
approach to weak equilibrium in progress.  The so called ``back
resonances'' in beta decay proceed through the same Gamow-Teller
resonance that dominates in electron capture.

Not every reaction is so balanced though.  The flow due to
$^{61}$Co(e$^- \bar \nu)^{61}$Ni at the same conditions is about
twelve times smaller than that from $^{61}$Ni(e$^-,\nu)^{61}$Co.

\ifthenelse{\boolean{emul}}{
\vspace{1.5\baselineskip}
\noindent
\includegraphics[width=\columnwidth]{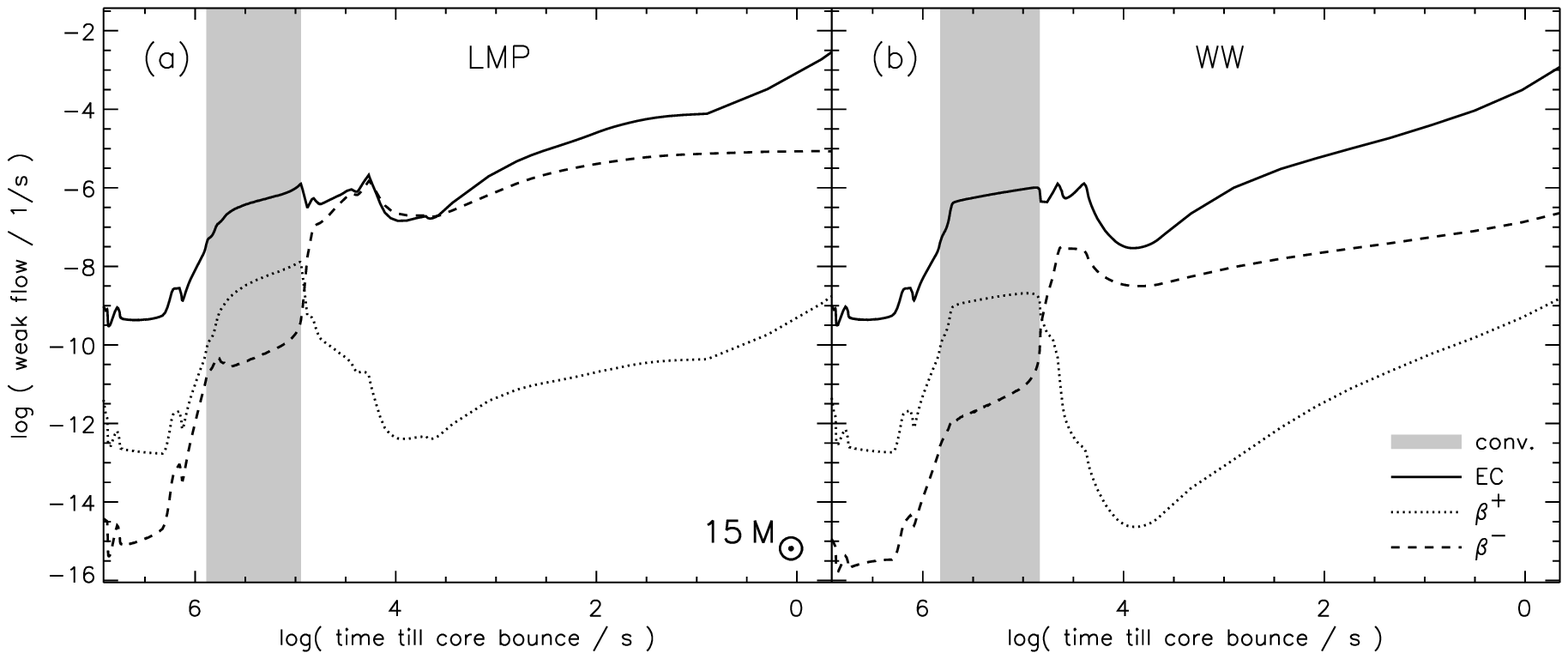}
\figcaption{\FigSix}
\vspace{1.5\baselineskip}}{}

\ifthenelse{\boolean{emul}}{
\vspace{1.5\baselineskip}
\noindent
\includegraphics[width=\columnwidth]{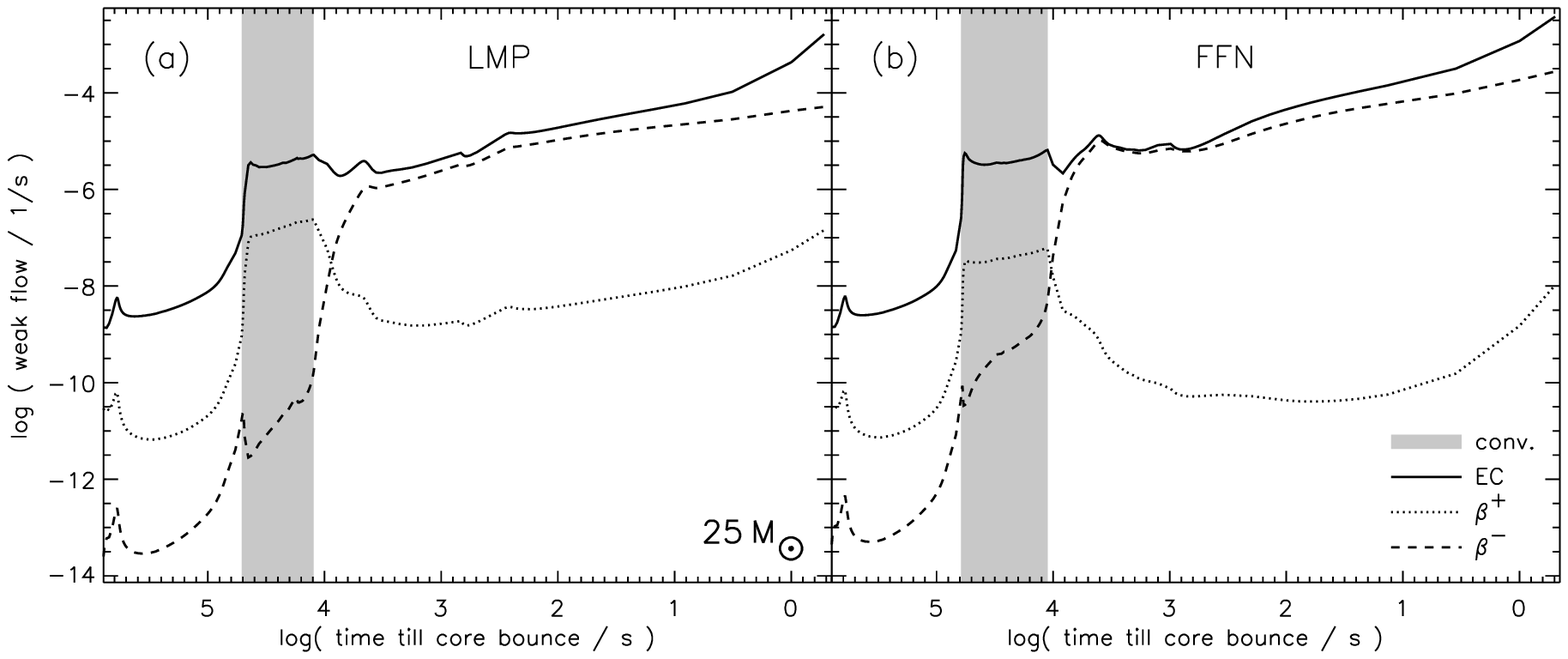}
\figcaption{\FigSeven}
\vspace{1.5\baselineskip}}{}

\ifthenelse{\boolean{emul}}{
\vspace{1.5\baselineskip}
\noindent
\includegraphics[width=\columnwidth]{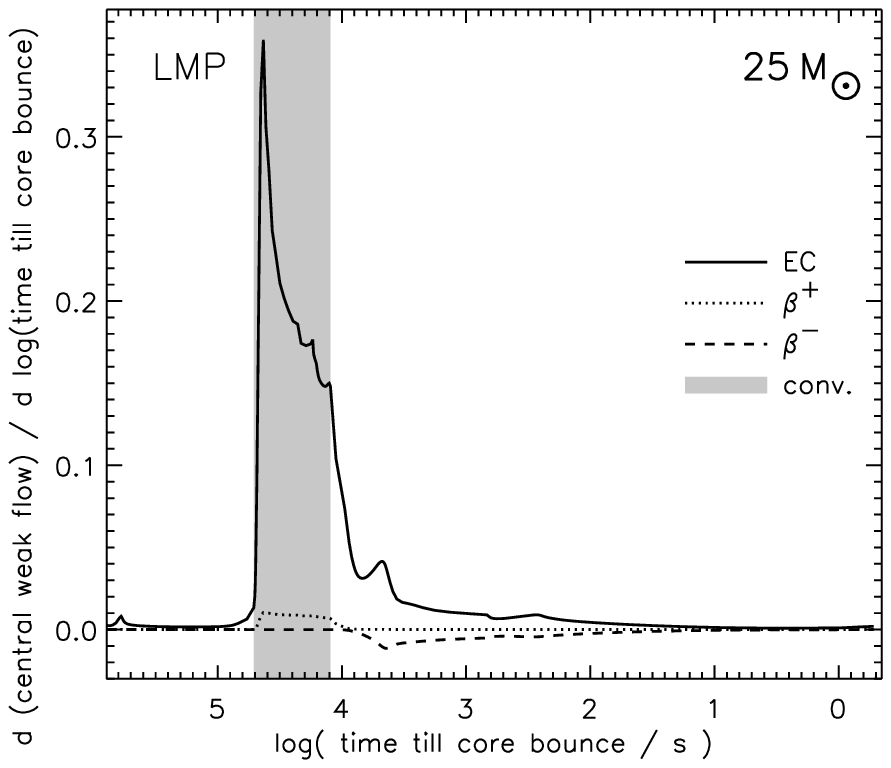}
\figcaption{\FigEight}
\vspace{1.5\baselineskip}}{}

\subsection{Sensitivity to Beta-Rates and Network}
\lSect{betasens}

Because of the state of near beta-equilibrium, the actual value of the
beta-decay rates does not greatly matter, so long as they are large
enough to instigate the equilibrium. Calculations in which all
beta-decay rates are arbitrarily multiplied by factors of 2 and 5 give
very similar results to those in which the standard values are used
(\Fig{yeoft15} and \Fig{yeoft25}; plots with beta-decay rates times 5,
omitted for clarity in those figures, are very similar to those where
the rates were multiplied by 2).  However, if the beta rates become
sufficiently small (approximated by zero here), the evolution is
greatly altered and the results with the new rates resemble more
closely what was obtained in the WW models .

One might be concerned about the relative sparsity of the reaction
network used here to calculate abundances in nuclear equilibrium, 125
isotopes.  This network, very similar to that used by \cite{WZW78}, is
reasonably complete up to nickel, but contains no heavier isotopes. It
contains 67 iron-group species including $^{44-52}$Ti, $^{47-54}$V,
$^{48-56}$Cr, $^{51-58}$Mn, $^{52-62}$Fe, $^{54-64}$Co, and
$^{56-66}$Ni.  Studies by \cite{auf94b}, using a much larger network,
suggest that this group of isotopes should be adequate to represent
both electron capture and beta decay for $\Ye \gtaprx 0.43$ (see their
Tables 16, 17, 22, and 23).  The larger values of \Ye we now obtain
for presupernova stars (\Tabs{15} -- \Tabff{40}) mean that the current
network is more applicable. However, we are concerned that the
isotopes contributing most to beta decay in the presupernova model
frequently lie on the neutron-rich boundary of our network. We intend
to expand its size to include both more neutron-rich isotopes in the
iron group and more flows above Ni in the near future.

\subsection{Important individual weak rates}
\lSect{flows}

By far the most important electron-capture rate in calculations that
use the FFN rates is $^{60}$Co(e$^-$,$\nu$)$^{60}$Fe.  This dominates
the evolution of \Ye in the center of the star from silicon depletion
onwards.  This particular rate is greatly reduced in the new LMP rate
set \cite[]{LM99}, and this is one of the main reasons \Ye is
larger in the cores of the new stars (\Tabs{15} -- \Tabff{flow40};
\Figs{contrib15} and \Figff{contrib25}).

In the new models using LMP rates, several electron capture reactions
dominate (\Tabs{15} -- \Tabff{40}).  Even capture on free protons is
no longer negligible, particularly in more massive stars at late times
(\Figs{contrib15} and \Figff{contrib25}).

\ifthenelse{\boolean{emul}}{
\vspace{1.5\baselineskip}
\noindent
\includegraphics[width=\columnwidth]{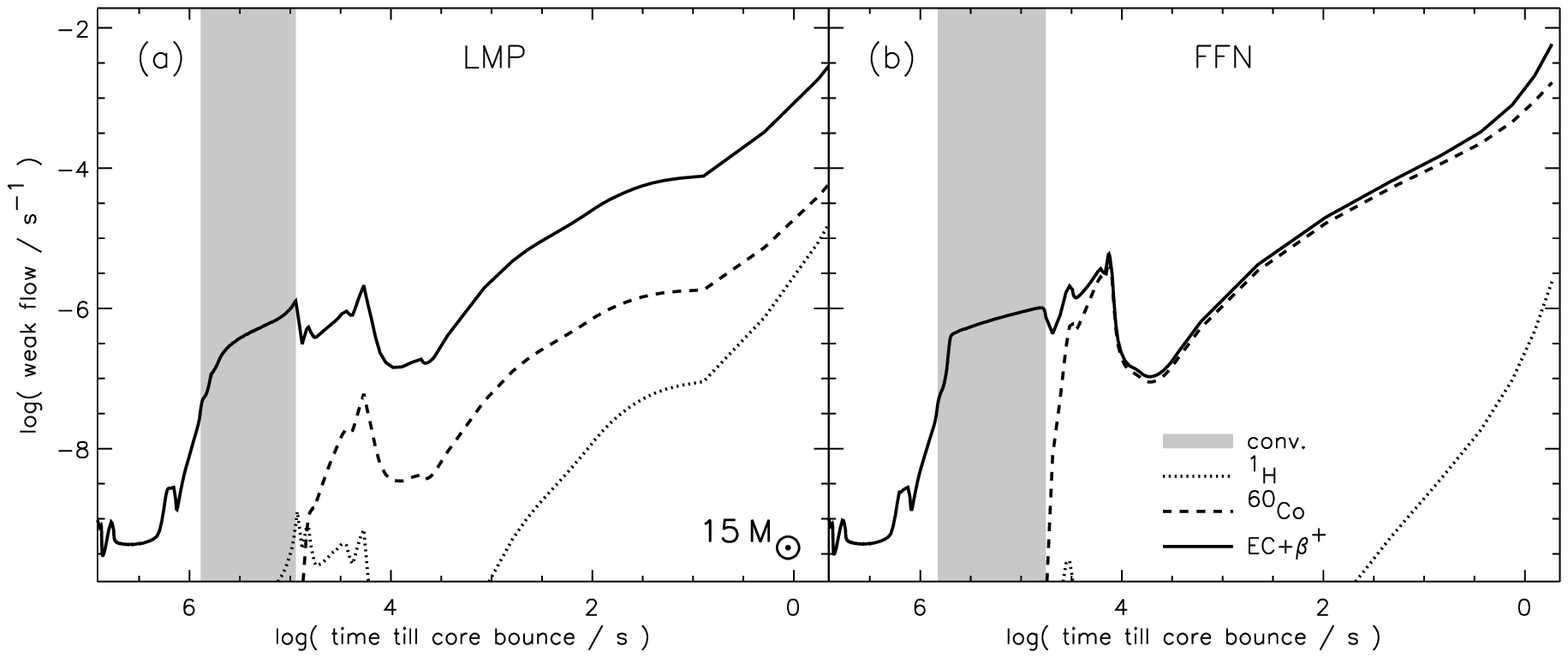}
\figcaption{\FigNine}
\vspace{1.5\baselineskip}}{}

\ifthenelse{\boolean{emul}}{
\vspace{1.5\baselineskip}
\noindent
\includegraphics[width=\columnwidth]{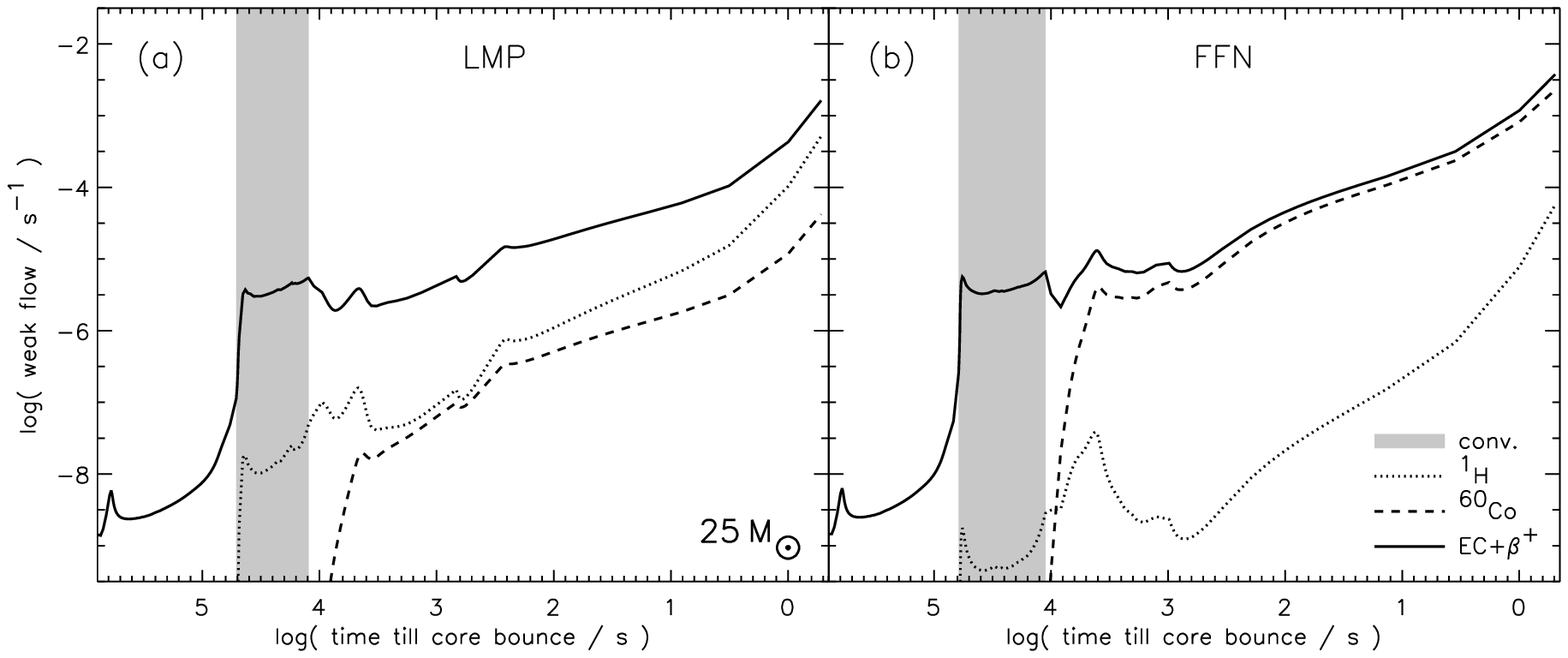}
\figcaption{\FigTen}
\vspace{1.5\baselineskip}}{}

\section{Changes in the Presupernova Structure}
\lSect{presn}

The structure of the presupernova star is altered both by the changes
in \Ye and entropy in its deep interior.  The altered composition and
entropy also affect the location and extension of convective shells
and can cause changes in structure that extend outside the region
where the rates themselves were altered. For some masses of star that
may have been on the verge of discontinuous behavior, e.g., igniting
an additional convective shell of silicon or oxygen burning, such
changes can be quite dramatic.

The general behavior of the density structure and iron core mass in
terms of electronic entropy per baryon, \Se, and \Ye can be
understood in terms of the definition of the Chandrasekhar mass
generalized for finite temperature \cite[][, e.g.]{Cha38,Timmes96}
\begin{equation}
M_\mathrm{Ch} \simeq  5.83 \; \Ye^2
\left [ 1 \ + \ \left(\frac{\Se}{\pi \Ye}\right)^2 \right]
\;.
\end{equation}
Here \Se is expressed in units of Boltzmann's constant, \kB, and is
dimensionless.  Note that, under circumstances in massive stars, the
entropy can be just as important as \Ye in determining the effective
Chandrasekhar Mass.

The practical difficulty applying this equation is that both \Ye and
\Se vary with the location in the core. \Figs{15presn} and
\Figff{25presn} show the composition, density, $\Ye(m)$, $\Se(m)$, and
total entropy $\St(m)$ in the cores of 15 and 25\,\Msun stars.

\ifthenelse{\boolean{emul}}{
\vspace{1.5\baselineskip}
\noindent
\includegraphics[width=\columnwidth]{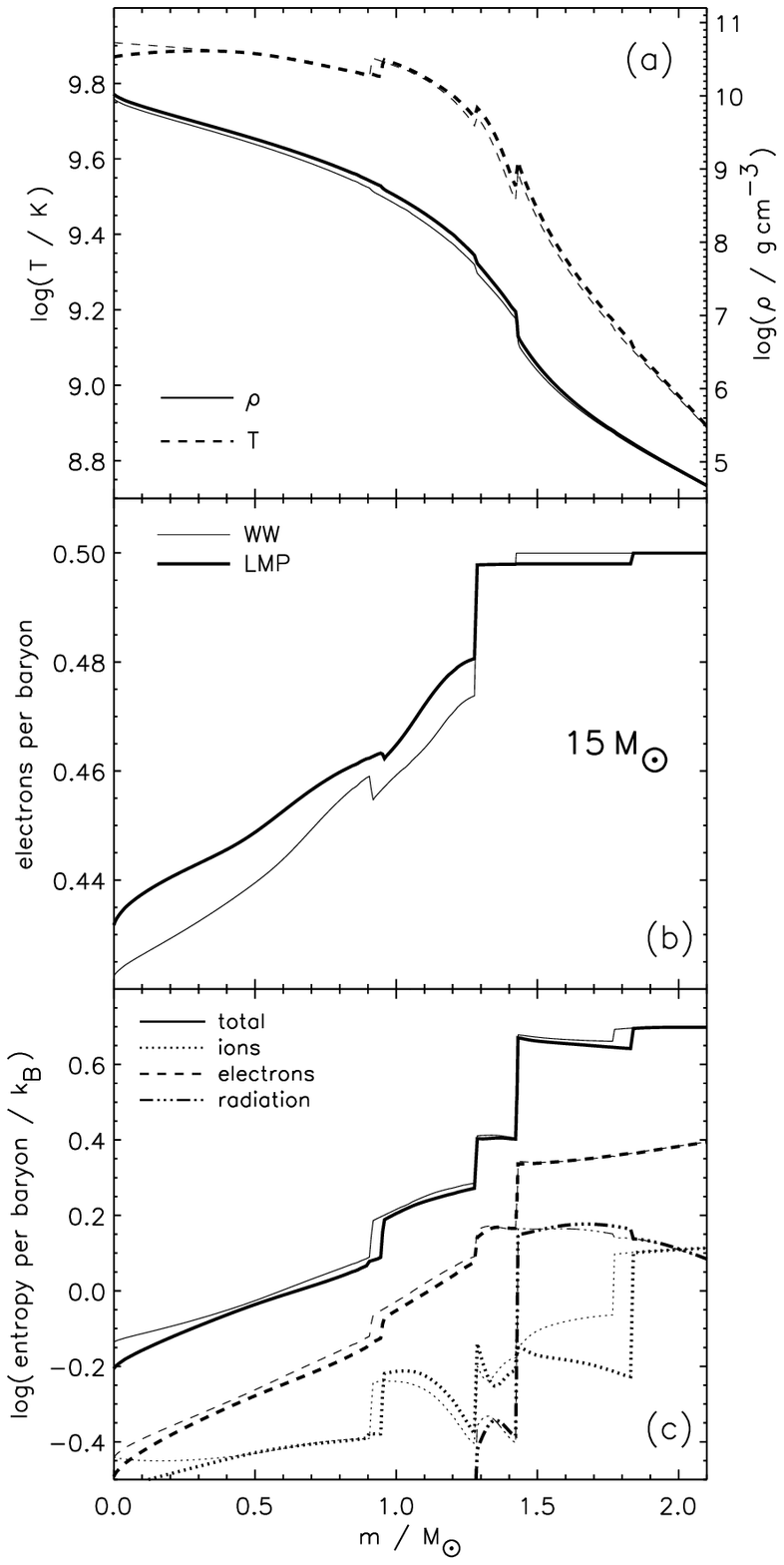}
\figcaption{\FigEleven}
\vspace{1.5\baselineskip}}{}

\ifthenelse{\boolean{emul}}{
\vspace{1.5\baselineskip}
\noindent
\includegraphics[width=\columnwidth]{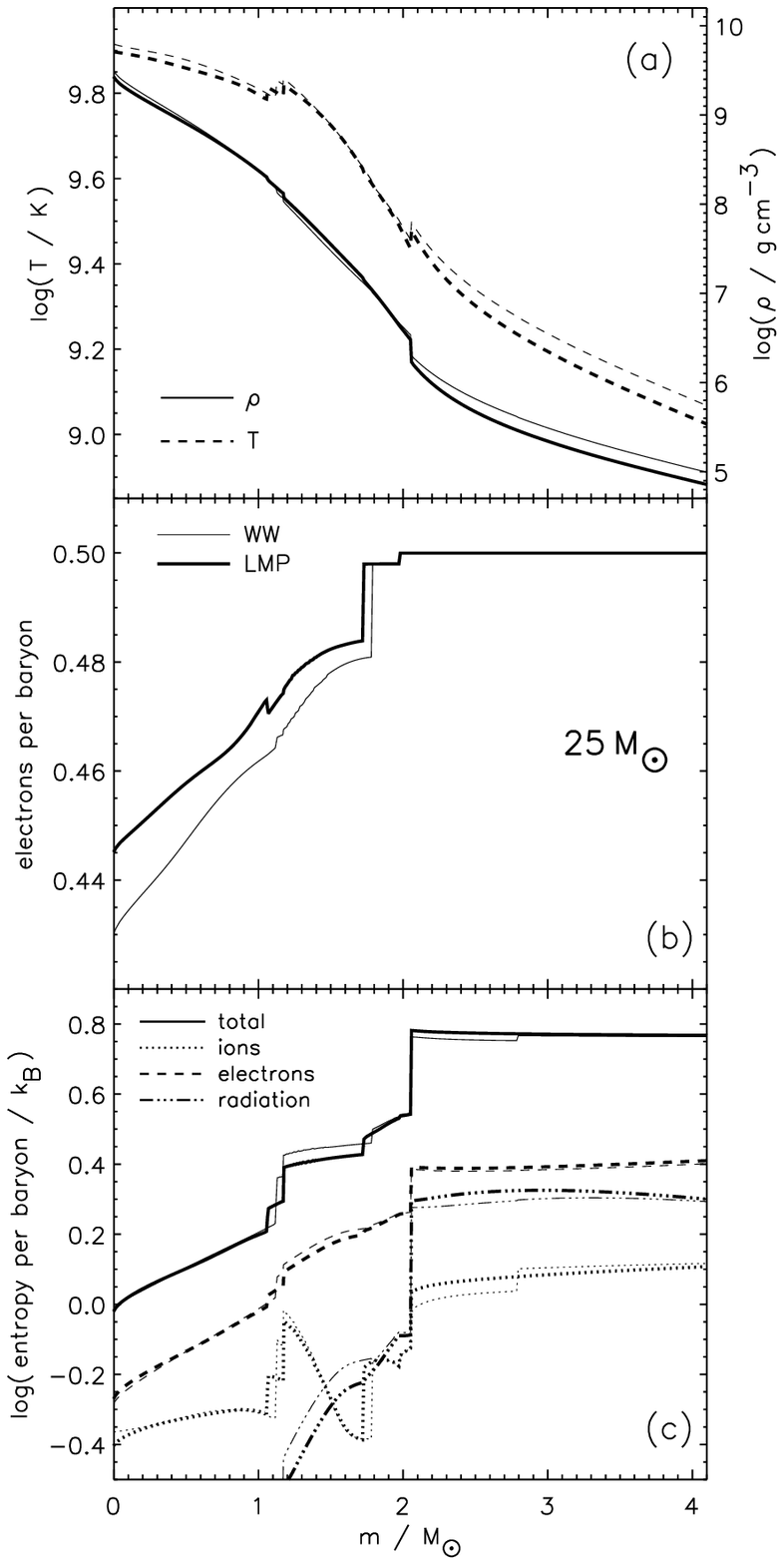}
\figcaption{\FigTwelve}
\vspace{1.5\baselineskip}}{}

\clearpage

A second difficulty is that the ``iron core'' and the core one
compares to the Chandrasekhar mass are not necessarily the
same.  Typically, unless there is explosive burning during the
collapse, the outer boundary of the iron core marks the final
extension of the last stage of silicon shell burning.  There is usually
a discontinuous jump of entropy at that point and a consequent fall
off in density.  This makes the iron core a meaningful parameter in the
calculation of the explosion mechanism. However, this jump in entropy
is not nearly so large as the one associated with the very vigorous
oxygen burning shell at the outer edge of the silicon shell.  Thus the
iron core is only approximately isolated from the overlying
material. Its surface boundary pressure is often not negligible.

Still, several general tendencies are evident. First, $\Ye(m)$ is
larger for the new rates \emph{everywhere} in all the new cores.  This
must act to increase the mass of the compact core (and the iron core
within) when it collapses. However, the entropy profile also shows
significant changes.  As \Fig{entropy} shows, the central entropy is
\emph{lower} for the new models (LMP) compared with old ones (WW) for
stellar masses up to about 27\,\Msun, and there is a tendency, at
least at low mass, for the decrease in entropy to correlate inversely
with the mass. Inclusion of beta-decay, which was left out by WW,
leads to weak equilibrium in the centers of stars in this mass range
(\Sect{betaequil}; \Figs{lmp15} and \Figff{lmp25}) and the
accompanying extra neutrino losses from this URCA-like process gives a
decreased entropy.  This decrease in entropy helps to offset the
increase in \Ye in determining the iron core mass. However, above
about 27\,\Msun the central entropies are \emph{higher}. Full beta
equilibrium is not achieved in these stars (using the LMP rates) and
the neutrino cooling that would have accompanied such equilibrium is
decreased. Moreover, there is less electron capture with the new
rates, and less neutrino cooling by this process as well, so the
entropy is higher. If central entropy were all that mattered, the
larger \Ye in these big stars would imply larger iron cores, contrary
to what is observed (\Fig{yefeofm}).

\ifthenelse{\boolean{emul}}{
\vspace{1.5\baselineskip}
\noindent
\includegraphics[width=\columnwidth]{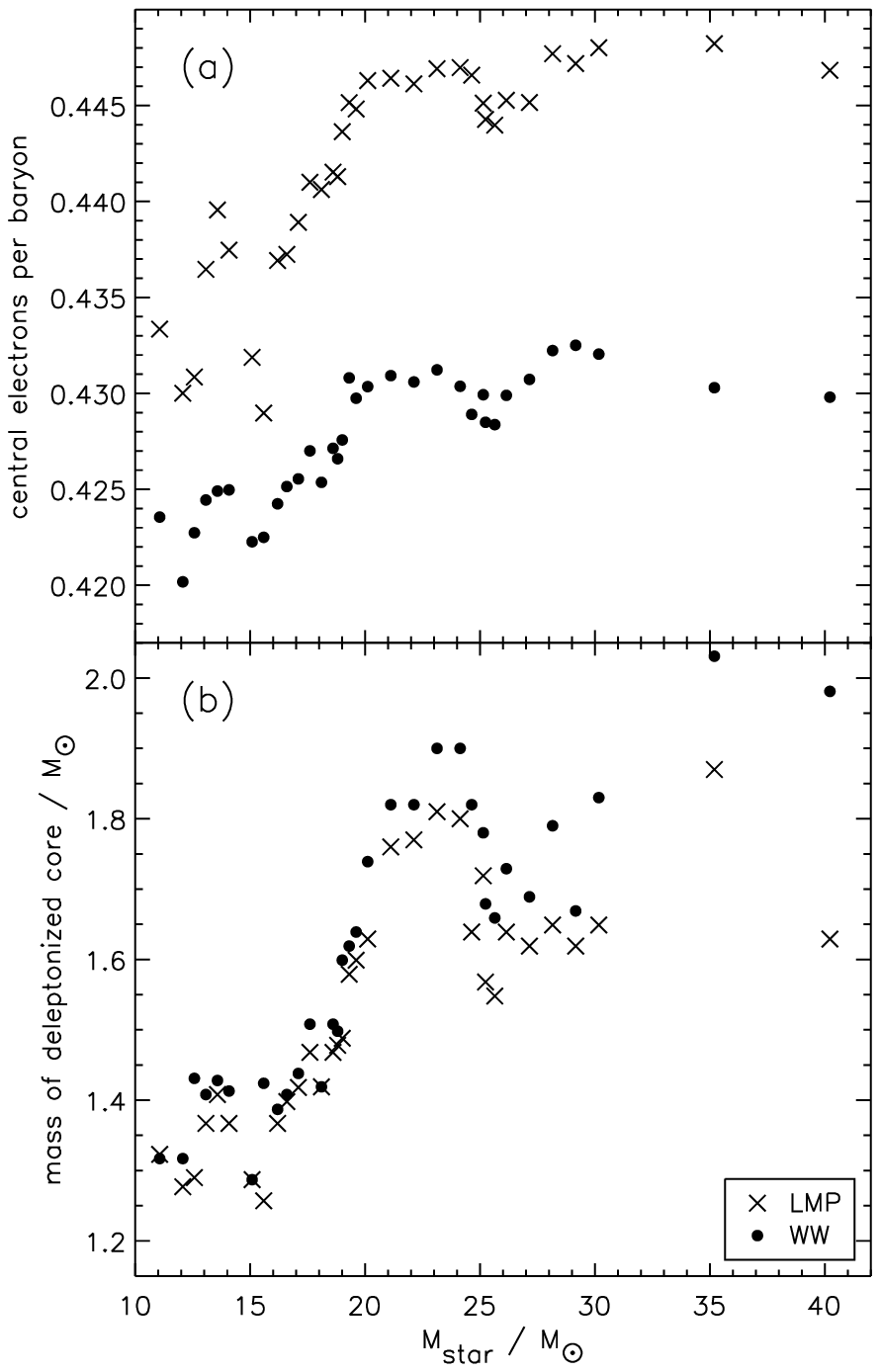}
\figcaption{\FigThirteen}
\vspace{1.5\baselineskip}}{}

\ifthenelse{\boolean{emul}}{
\vspace{1.5\baselineskip}
\noindent
\includegraphics[width=\columnwidth]{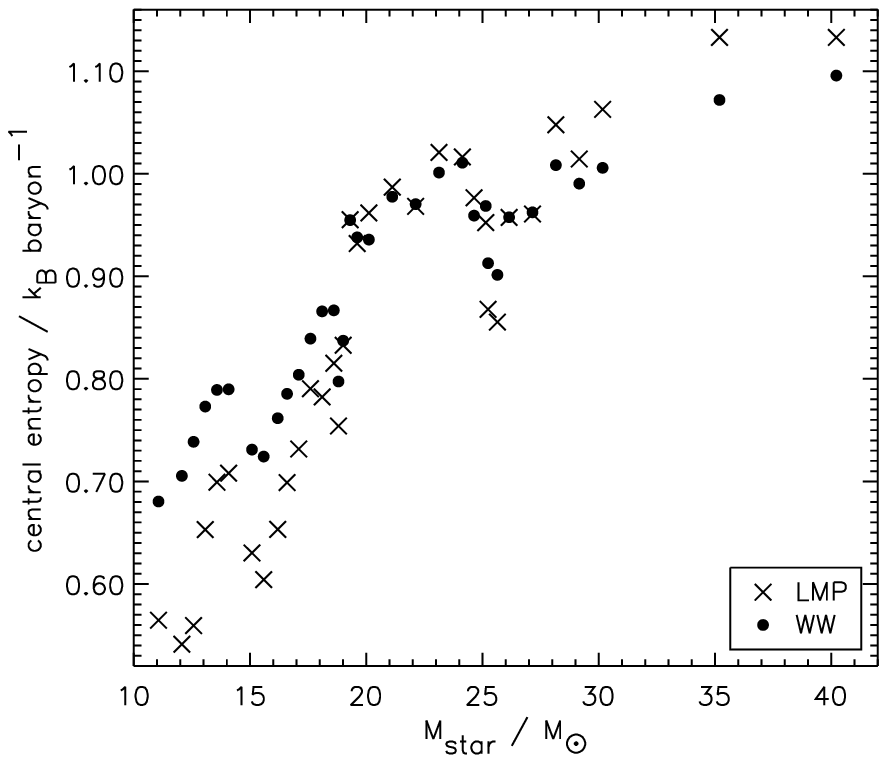}
\figcaption{\FigFourteen}
\vspace{1.5\baselineskip}}{}
	
\ifthenelse{\boolean{emul}}{
\vspace{1.5\baselineskip}
\noindent
\includegraphics[width=\columnwidth]{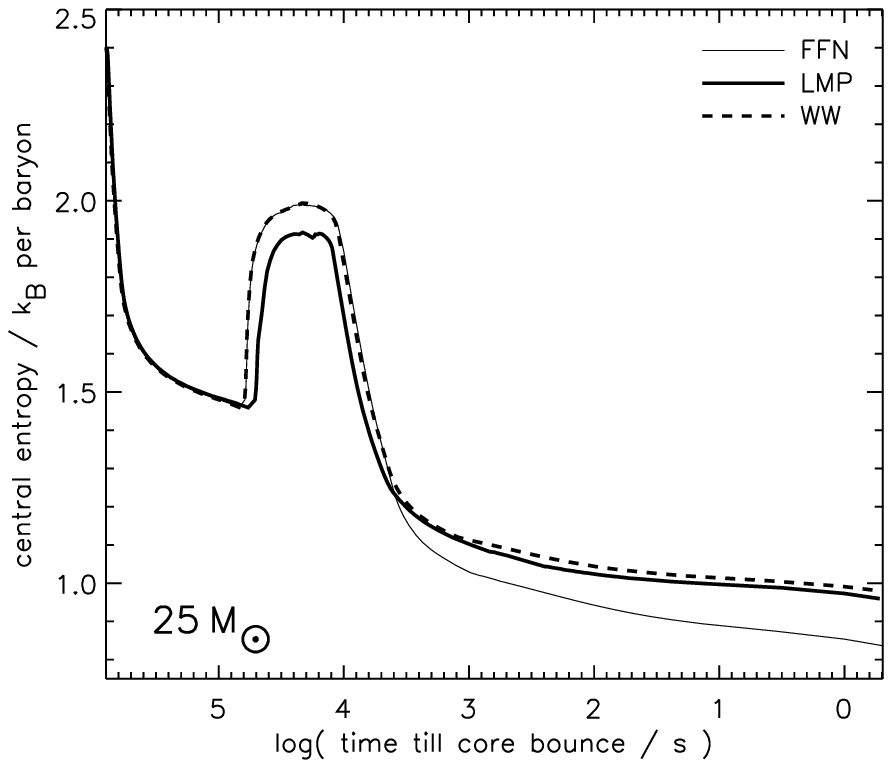}
\figcaption{\FigFifteen}
\vspace{1.5\baselineskip}}{}

This seeming paradox results because conditions in the outer core are
more important to determining the final distribution of composition
than those in the inner core. None of our stars collapse following the
first stage of silicon core burning. All have one or more stages
during which silicon burns in a convective shell around an inert iron
core. It is the extent of these convective shells that determines the
boundary between iron- and silicon-rich compositions in the
pre-supernova star.  The extent of these convective shells is sensitive
to the entropy in the surrounding material, the entropy developed in
the convective shell, and its composition. 

For convection according to the Ledoux criterion, as used in KEPLER,
the change in \Ye between two zones affects the criteria for mixing.
It is more difficult to mix two regions if the \Ye of the underlying
material is smaller.  However, the difference is not great and it
actually acts in the opposite direction to what is observed.  That is,
the smaller \Ye in the FFN models should make the growth of the
convective shell more difficult and the iron core smaller. Clearly
another effect is at work - the entropy itself.

The entropy in the convective silicon shell (and during silicon core
burning; \Tabs{15} -- \Tabff{40}) is smaller with the new rates
(\Figs{15presn} and \Figff{25presn}).  During silicon burning, beta
decay has not yet become important (\Figs{lmp15} and \Figff{lmp25}).
Differences that arise in the silicon shell between the old WW models
and the new ones are thus solely a consequence of the altered rates
for and neutrino losses associated with electron capture and not
effected by the neglect, by WW, of beta decay.  The larger value of
entropy obtained with the old rates is partly a consequence of
increased energy generation during silicon burning (\Tabs{15} --
\Tabff{40}, see ``Si core burning'').  Lowering \Ye (i.e., using the
FFN rates) means that silicon burns to nuclei in the iron group that
are more tightly bound.  $^{56}$Fe is more tightly bound than
$^{54}$Fe which is, in turn, more tightly bound than $^{56}$Ni.
Neutronization also brings into play more efficient channels of
silicon destruction.  For \Ye near 0.50, silicon burns at a rate set
by $^{24}$Mg($\gamma,\alpha)^{20}$Ne \cite[]{bod68}.  For \Ye
appreciably below 0.50 though, $^{26}$Mg(p,$\alpha)^{23}$Na becomes an
efficient alternative using $^{26}$Mg that is in quasi-equilibrium
with $^{28,29,30}$Si.  Consequently, silicon burning is a little
briefer and yields a little more energy per gram with the FFN
rates. Because of the higher specific energy, it can burn at a lower
temperature. Because of this and the shorter duration, there are less
neutrino losses to pair processes.  The mean neutrino energies are
also larger when one captures on nuclei that are abundant at larger
\Ye.  For all these reasons, the entropy associated with silicon
burning is a little bit less with the new rates.

The effects on the iron core mass can be cumulative. Because the
entropy of the first stage of convective core burning is less
(\Fig{entropyoft}), the 25\,\Msun star ends up igniting its silicon
shell at 1.07\,\Msun instead of 1.12\,\Msun \ (\Fig{25presn}).  This
puts the silicon shell deeper in the ``entropy well'' to begin
with.  Then the entropy of shell burning is also lower.  Combining
these effects leads to the silicon convective shell not extending as
far as it did with the old rates.  Because the entropy gradients,
and the consequent barrier to the growth of the convective shell, are
shallower in more massive stars, the change in the iron core mass is
larger there.

\subsection{Implications for the Explosion Mechanism and Neutron Star Masses}
\lSect{nstar}

Despite the emphasis placed upon differences in \Ye and entropy in the
previous section, the most striking feature of the density plots in
\Fig{15presn} and \Fig{25presn} is their similarity.  The appreciable
changes in iron core mass are not really accompanied by large changes
in the core structure.  The location of the large entropy jump
associated with the oxygen burning shell also does not change
(\Fig{sicore}). Still, the success or failure of explosion
calculations often hinges on small differences in the accretion rate,
radius of the stalled shock, and neutrino luminosity
\cite[e.g.,][]{jan00}.  One may expect revisions, if only in the net
explosion energy.  These revisions cannot be anticipated in advance
because they rely on the highly non-linear outcome of a system where
more than one key parameter is being varied.

\ifthenelse{\boolean{emul}}{
\vspace{1.5\baselineskip}
\noindent
\includegraphics[width=\columnwidth]{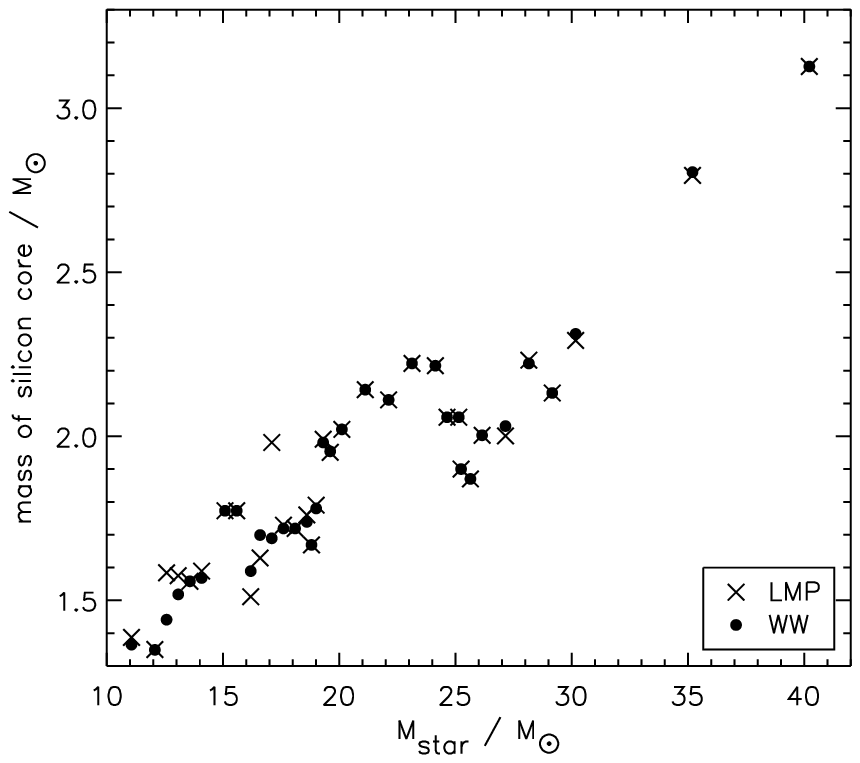}
\figcaption{\FigSixteen}
\vspace{1.5\baselineskip}}{}

Generally speaking, iron cores with larger \Ye and smaller mass are
easier to explode.  The larger \Ye implies a larger ``homologous
core'' and the prompt shock is born farther out with less overburden
\cite[]{Bethe90}.  A smaller iron core also decreases the losses the
shock experiences to photodisintegration before breaking free. But it
is very doubtful that the new models are so different as to allow
``prompt'' hydrodynamic explosions.  That being the case, the density
just outside the iron core is also very important as that sets the
``ram pressure'' that any neutrino-powered explosion must
overcome.  The density in this region is larger in the new models by as
much as 50\,\%. So, in the neutrino powered models, currently favored,
the new models could actually be more difficult to explode.

\section{Conclusions}
\lSect{conclude}

We have explored in some detail the weak interactions that
characterize the last few days in the life of a massive star. In
particular we have studied the effects of a new, improved set of weak
interaction rates and neutrino losses by \cite{LM00} and 
the role played by beta-decay as well as electron capture.

In agreement with \cite{auf94a}, we find that inclusion of beta decay
makes a significant difference in the presupernova structure. Beta
equilibrium comes to exist, temporarily, in the center of the star
after silicon is depleted.  This causes the final electron mole number,
\Ye, to be larger than in models where beta decay has been
neglected \cite[e.g.,][]{WW95}.  The existence of beta equilibrium,
particularly in the lower mass stars ($M \ltaprx 20\,\Msun$), also
decreases the central entropy. Both effects alter the structure of the
presupernova star.

In addition to the consequences of including beta decay, we find that
which rate set one uses also makes a difference. In particular, for
the more massive stars studied here, use of the new weak rates of LMP
instead of the older set from FFN causes an additional increase in
\Ye comparable to that from the inclusion of beta decay.  This is
because the new rates for electron capture are systematically smaller
owing to a more accurate treatment of the Gamow-Teller resonance. All
in all, \Ye in the iron cores of the new models is systematically
larger by approximately 3\,\% (\Fig{yefeofm}). Since \Ye enters
quadratically into determining the Chandrasekhar Mass and since the
mass of the iron core is critically important in models for the
supernova explosion mechanism, this is a significant difference.

However, contrary to simple expectations based only upon the change in
\Ye, the new iron core masses are systematically about 0.05 to
0.1\,\Msun \emph{smaller} than in WW (\Fig{yefeofm}).  This has
several causes, but chiefly reflects the lower entropy in the
convective silicon shell just prior to iron core collapse.  The lower
entropy inhibits the final growth of the iron core in a surrounding
medium of fixed entropy.

In total, the new presupernova models may explode more easier due both
to their reduced iron core mass and larger \Ye, however the changes
are not large and their are other effects (lower entropy just outside
the iron core) that could act in the opposite direction.  The new 15,
20, and 25\,\Msun models discussed here are available to those
wanting to study the explosion mechanism.

As \Tabs{15} -- \Tabff{flow40} show, some of the most interesting weak
processes in presupernova stars involve unstable, radioactive nuclei
for which no experimental data exists about the GT strength
distribution.  This situation should change once the proposed
Rare-Isotope Accelerator (RIA) and similar facilities become
operational.  These accelerators will allow studies of the GT
distributions in unstable nuclei by charge-changing reactions like
(d,$^2$He) or ($^{12}$C,$^{12}$N) in inverse kinematics.  Such
experiments, as well as high-precision beta-decay measurements will
either verify or severely constrain the nuclear models.  They will
also determine energies, quantum numbers, and low-lying transition
strengths at low excitation energies.

For example, the shell model calculations predict that, with
increasing neutron excess, the centroid of the GT distribution moves
to lower excitation energies in the daughter nucleus \cite[see Fig.~8
in]{LM00}.  It would be very
helpful to quantify this prediction experimentally. Such data could
appreciably improve our understanding of proton-neutron correlations
and of the evolution of the $f_{7/2}-f_{5/2}$ spin-orbit splitting in
the $pf$-shell.

There are also currently no experimental data regarding the GT$_+$
distribution for odd-odd nuclei in the $pf$-shell.  This is
particularly unfortunate given the key role played by capture on
$^{60}$Co in the FFN rates.  The shell model electron capture rates
for odd-odd nuclei, however, are significantly -- often by more than
one order of magnitude at the relevant temperature and density -- lower
than the FFN estimates and this is largely responsible for the
difference between the FFN and LMP presupernova models.  This is
because the shell model places the centroid of the GT strength
distribution for odd-odd nuclei at noticeably higher excitation
energies in the daughter nucleus.

\acknowledgements We are grateful to Rob Hoffman, for help in
implementing the different versions of the weak rates into the KEPLER
code, and to Tom Weaver for providing the stellar models starting at
oxygen depletion that were used for the finely binned (in mass) study
of stellar models and the presupernova models that used the input
physics of WW.  This work was supported by the National Science
Foundation (AST-97-31569 and AST-INT-9909999), by the ASCI Program of
the U.S.\ Department of Energy by under contract No.\ W-7405-Eng-48,
by the Alexander von Humboldt-Stiftung (FLF-1065004), and by the
Danish Research Council.  GMP was supported by the Carlsberg Foundation.

{}

\newpage

{
\renewcommand{\E}[1]{&{\ensuremath{\cdot10^{#1}}}}
\newcommand{\EE}{&}
\renewcommand{\I}[2]{{\ensuremath{^{#1}}}&{\ensuremath{\mathrm{#2}}}}

\begin{table}
\ifthenelse{\boolean{emul}}{}{\tiny}
\caption{Properties of the stellar center of a 15\,\Msun star \lTab{15}}
\begin{tabular}{l|r@{}lr@{}l|r@{}lr@{}l|r@{}lr@{}l|r@{}lr@{}l}
\hline
\hline
& \multicolumn{4}{c|}{O-dep}
& \multicolumn{4}{c|}{O-shell}
& \multicolumn{4}{c|}{Si-ign}
& \multicolumn{4}{c}{Si-burn} \\
& \multicolumn{2}{c}{FFN} & \multicolumn{2}{c|}{LMP}
& \multicolumn{2}{c}{FFN} & \multicolumn{2}{c|}{LMP}
& \multicolumn{2}{c}{FFN} & \multicolumn{2}{c|}{LMP}
& \multicolumn{2}{c}{FFN} & \multicolumn{2}{c}{LMP} \\
\hline
 $\log(\tb-t)$ &       6.91 & &       6.91 & &       6.51 & &       6.51 & &       5.84 & &       5.90 & &       5.30 & &       5.30 & \\
           $T$ &  2.26\E{  9} &  2.26\E{  9} &  1.90\E{  9} &  1.90\E{  9} &  2.87\E{  9} &  2.86\E{  9} &  3.32\E{  9} &  3.39\E{  9} \\
        $\rho$ &  1.23\E{  7} &  1.24\E{  7} &  2.74\E{  7} &  2.75\E{  7} &  1.09\E{  8} &  1.08\E{  8} &  3.88\E{  7} &  4.46\E{  7} \\
         $\Ye$ &  4.98\E{- 1} &  4.98\E{- 1} &  4.95\E{- 1} &  4.95\E{- 1} &  4.89\E{- 1} &  4.89\E{- 1} &  4.76\E{- 1} &  4.80\E{- 1} \\
&&&&&&&&&&&&\\
        \snucq &  9.63\E{ 10} &  9.24\E{ 10} &  1.01\E{  9} &  1.01\E{  9} &  1.36\E{ 11} &  1.08\E{ 11} &  3.28\E{ 13} &  2.97\E{ 13} \\
         \snuw &  1.22\E{  9} &  1.19\E{  9} &  4.23\E{  8} &  4.21\E{  8} &  4.38\E{ 10} &  3.92\E{ 10} &  1.12\E{ 12} &  9.08\E{ 11} \\
       \snubps &  1.52\E{ 11} &  1.52\E{ 11} &  3.76\E{  9} &  3.81\E{  9} &  6.49\E{ 10} &  6.24\E{ 10} &  2.45\E{ 12} &  2.56\E{ 12} \\
        \sdotq & -5.74\E{ 10} & -6.10\E{ 10} & -3.18\E{  9} & -3.21\E{  9} &  2.74\E{ 10} &  6.13\E{  9} &  2.92\E{ 13} &  2.62\E{ 13} \\
&&&&&&&&&&&&\\
          \sig &  1.78\EE     &  1.78\EE     &  1.34\EE     &  1.34\EE     &  1.25\EE     &  1.25\EE     &  1.57\EE     &  1.52\EE     \\
         \sige &  1.03\EE     &  1.03\EE     &  6.77\E{- 1} &  6.77\E{- 1} &  6.20\E{- 1} &  6.20\E{- 1} &  9.64\E{- 1} &  9.47\E{- 1} \\
         \sigi &  6.35\E{- 1} &  6.36\E{- 1} &  6.28\E{- 1} &  6.28\E{- 1} &  6.02\E{- 1} &  6.03\E{- 1} &  4.88\E{- 1} &  4.68\E{- 1} \\
         \sigr &  1.13\E{- 1} &  1.13\E{- 1} &  3.03\E{- 2} &  3.03\E{- 2} &  2.65\E{- 2} &  2.64\E{- 2} &  1.14\E{- 1} &  1.06\E{- 1} \\
         \sigp &  7.55\E{- 4} &  7.53\E{- 4} &  1.11\E{- 5} &  1.11\E{- 5} &  1.76\E{- 5} &  1.72\E{- 5} &  1.69\E{- 3} &  1.43\E{- 3} \\
&&&&&&&&&&&&\\
           \Yp &  1.31\E{-13} &  1.29\E{-13} &  4.85\E{-17} &  5.18\E{-17} &  1.67\E{-12} &  1.53\E{-12} &  2.16\E{- 9} &  1.40\E{- 8} \\
        \rtotp &  1.97\E{-17} &  1.95\E{-17} &  1.88\E{-20} &  2.03\E{-20} &  3.04\E{-14} &  2.73\E{-14} &  7.26\E{-12} &  6.39\E{-11} \\
&&&&&&&&&&&&\\
        \wrate &  1.00\E{- 9} &  9.56\E{-10} &  4.40\E{-10} &  4.40\E{-10} &  3.04\E{- 8} &  2.72\E{- 8} &  6.76\E{- 7} &  5.26\E{- 7} \\
\hline
\rate(\EC+\pd) &  \I{ 35}{Cl} &  \I{ 35}{Cl} &  \I{ 35}{Cl} &  \I{ 35}{Cl} &  \I{ 33}{ S} &  \I{ 33}{ S} &  \I{ 53}{Mn} &  \I{ 54}{Fe} \\
               &  2.96\E{-10} &  2.96\E{-10} &  2.28\E{-10} &  2.34\E{-10} &  1.42\E{- 8} &  1.23\E{- 8} &  2.22\E{- 7} &  1.54\E{- 7} \\
               &  \I{ 37}{Ar} &  \I{ 37}{Ar} &  \I{ 33}{ S} &  \I{ 33}{ S} &  \I{ 35}{Cl} &  \I{ 35}{Cl} &  \I{ 55}{Fe} &  \I{ 55}{Fe} \\
               &  2.55\E{-10} &  2.55\E{-10} &  1.12\E{-10} &  1.12\E{-10} &  6.35\E{- 9} &  6.13\E{- 9} &  2.12\E{- 7} &  1.28\E{- 7} \\
               &  \I{ 32}{ S} &  \I{ 32}{ S} &  \I{ 37}{Ar} &  \I{ 37}{Ar} &  \I{ 31}{ P} &  \I{ 31}{ P} &  \I{ 54}{Fe} &  \I{ 53}{Mn} \\
               &  1.57\E{-10} &  1.43\E{-10} &  6.99\E{-11} &  7.09\E{-11} &  4.42\E{- 9} &  5.07\E{- 9} &  1.05\E{- 7} &  5.46\E{- 8} \\
               &  \I{ 36}{Ar} &  \I{ 36}{Ar} &  \I{ 32}{ S} &  \I{ 32}{ S} &  \I{ 32}{ S} &  \I{ 32}{ S} &  \I{ 57}{Co} &  \I{ 50}{Cr} \\
               &  1.06\E{-10} &  1.23\E{-10} &  1.25\E{-11} &  1.21\E{-11} &  2.74\E{- 9} &  2.23\E{- 9} &  3.31\E{- 8} &  3.86\E{- 8} \\
               &  \I{ 33}{ S} &  \I{ 33}{ S} &  \I{ 36}{Ar} &  \I{ 36}{Ar} &  \I{ 37}{Ar} &  \I{ 37}{Ar} &  \I{ 51}{Cr} &  \I{ 51}{Cr} \\
               &  9.90\E{-11} &  9.55\E{-11} &  5.08\E{-12} &  4.82\E{-12} &  7.51\E{-10} &  6.67\E{-10} &  2.52\E{- 8} &  3.20\E{- 8} \\
&&&&&&&&&&&&\\
       \rectot &  1.00\E{- 9} &  9.52\E{-10} &  4.39\E{-10} &  4.39\E{-10} &  3.03\E{- 8} &  2.72\E{- 8} &  6.75\E{- 7} &  5.21\E{- 7} \\
       \rpdtot &  4.46\E{-12} &  3.95\E{-12} &  1.98\E{-13} &  1.85\E{-13} &  6.05\E{-11} &  5.22\E{-11} &  1.58\E{- 9} &  5.49\E{- 9} \\
&&&&&&&&&&&&\\
\rate(\PC+\ed) &  \I{ 32}{ P} &  \I{ 32}{ P} &  \I{ 32}{ P} &  \I{ 32}{ P} &  \I{ 32}{ P} &  \I{ 32}{ P} &  \I{ 55}{Mn} &  \I{ 54}{Mn} \\
               &  5.08\E{-15} &  2.99\E{-15} &  1.77\E{-15} &  1.06\E{-15} &  1.92\E{-11} &  9.73\E{-12} &  5.75\E{-10} &  4.08\E{-11} \\
               &  \I{ 36}{Cl} &  \I{ 36}{Cl} &  \I{ 36}{Cl} &  \I{ 36}{Cl} &  \I{ 28}{Al} &  \I{ 28}{Al} &  \I{ 53}{Cr} &  \I{ 55}{Mn} \\
               &  8.68\E{-16} &  7.02\E{-16} &  7.70\E{-17} &  5.35\E{-17} &  5.70\E{-13} &  3.88\E{-13} &  1.29\E{-10} &  4.49\E{-12} \\
               &  \I{ 46}{Sc} &  \I{ 46}{Sc} &  \I{ 46}{Sc} &  \I{ 46}{Sc} &  \I{ 33}{ P} &  \I{ 33}{ P} &  \I{ 32}{ P} &  \I{ 32}{ P} \\
               &  2.65\E{-16} &  3.65\E{-17} &  2.80\E{-17} &  3.53\E{-18} &  4.48\E{-13} &  8.43\E{-14} &  4.12\E{-11} &  4.15\E{-12} \\
               &  \I{ 55}{Mn} &  \I{ 54}{Mn} &  \I{ 33}{ P} &  \I{ 54}{Mn} &  \I{ 55}{Mn} &  \I{ 52}{ V} &  \I{ 57}{Fe} &  \I{ 58}{Co} \\
               &  2.49\E{-17} &  2.26\E{-17} &  1.00\E{-17} &  2.85\E{-18} &  2.56\E{-13} &  4.53\E{-14} &  3.28\E{-11} &  2.64\E{-12} \\
               &  \I{ 33}{ P} &  \I{ 39}{Ar} &  \I{ 55}{Mn} &  \I{ 33}{ P} &  \I{ 53}{Cr} &  \I{ 36}{Cl} &  \I{ 54}{Cr} &  \I{ 50}{ V} \\
               &  2.00\E{-17} &  1.40\E{-17} &  5.96\E{-18} &  2.83\E{-18} &  9.41\E{-14} &  4.12\E{-14} &  3.05\E{-11} &  2.37\E{-12} \\
&&&&&&&&&&&&\\
       \redtot &  6.30\E{-15} &  3.78\E{-15} &  1.89\E{-15} &  1.12\E{-15} &  2.07\E{-11} &  1.03\E{-11} &  8.55\E{-10} &  6.06\E{-11} \\
\hline
\end{tabular}
\\[0.5\baselineskip] \textsc{Note.}---
For different stages of late stellar evolution (logarithm of time till
core collapse, $\tb-t$ in \Sec) we give for the FFN and LMP models the
central values of temperature ($T$ in \K), density ($\rho$ in \gcc),
electrons per baryon (\Ye), specific energy generation rates: nuclear
(\snucq), weak neutrino losses (\snuw), plasma neutrino losses
(\snubps), and total (\sdotq), all in \erggs, total entropy (\sig),
electron entropy (\sige), ion entropy (\sigi), radiation entropy
(\sigr) and entropy due to pairs (\sigp), all in \Sunit, the number
fraction of protons (\Yp), the rate of electron captures on protons
(\rtotp in \Seci), and the total net weak rate ($\wrate=$d$\Ye/$d$t$
in \Seci).  The different contribution to \wrate are shown separately
in the second section: we give the weak rates of the five most
important isotopes for \EC and \pd-decay (upper part) and for \PC and
\ed-decay (lower part), and the total rates of \EC (\rectot),
\pd-decay (\rpdtot), and \ed-decay (\redtot), all in \Seci per
nucleon.
\begin{flushright}\textsc{(continued on next page)}\end{flushright}
\end{table}

\clearpage

\addtocounter{table}{-1}

\begin{table}
\ifthenelse{\boolean{emul}}{}{\tiny}
\caption{(continued) Properties of the stellar center of a 15\,\Msun star}
\begin{tabular}{l|r@{}lr@{}l|r@{}lr@{}l|r@{}lr@{}l|r@{}lr@{}l}
\hline
\hline
& \multicolumn{4}{c|}{Si-dep}
& \multicolumn{4}{c|}{Si-shell}
& \multicolumn{4}{c|}{core-contr}
& \multicolumn{4}{c}{pre-SN} \\
& \multicolumn{2}{c}{FFN} & \multicolumn{2}{c|}{LMP}
& \multicolumn{2}{c}{FFN} & \multicolumn{2}{c|}{LMP}
& \multicolumn{2}{c}{FFN} & \multicolumn{2}{c|}{LMP}
& \multicolumn{2}{c}{FFN} & \multicolumn{2}{c}{LMP} \\
\hline
 $\log(\tb-t)$ &       4.75 & &       4.93 & &       4.50 & &       4.53 & &       3.51 & &       3.57 & &      -0.27 & &      -0.60 & \\
           $T$ &  3.59\E{  9} &  3.78\E{  9} &  4.00\E{  9} &  4.13\E{  9} &  3.12\E{  9} &  3.55\E{  9} &  6.18\E{  9} &  7.16\E{  9} \\
        $\rho$ &  4.21\E{  7} &  5.87\E{  7} &  1.74\E{  8} &  3.20\E{  8} &  4.79\E{  8} &  5.41\E{  8} &  7.23\E{  9} &  9.08\E{  9} \\
         $\Ye$ &  4.63\E{- 1} &  4.67\E{- 1} &  4.52\E{- 1} &  4.49\E{- 1} &  4.42\E{- 1} &  4.45\E{- 1} &  4.30\E{- 1} &  4.32\E{- 1} \\
&&&&&&&&&&&&\\
        \snucq &  3.49\E{ 12} &  2.65\E{ 12} & -5.27\E{ 10} & -1.11\E{ 11} &  1.91\E{ 11} & -1.16\E{ 10} & -1.44\E{ 16} & -2.16\E{ 16} \\
         \snuw &  1.30\E{ 12} &  1.73\E{ 12} &  4.94\E{ 12} &  1.91\E{ 12} &  5.23\E{ 11} &  6.06\E{ 11} &  1.72\E{ 16} &  2.46\E{ 16} \\
       \snubps &  5.20\E{ 12} &  5.60\E{ 12} &  1.82\E{ 12} &  7.87\E{ 11} &  7.64\E{  9} &  2.96\E{ 10} &  3.00\E{ 11} &  7.04\E{ 11} \\
        \sdotq & -3.00\E{ 12} & -4.68\E{ 12} & -6.81\E{ 12} & -2.80\E{ 12} & -3.39\E{ 11} & -6.47\E{ 11} & -3.16\E{ 16} & -4.61\E{ 16} \\
&&&&&&&&&&&&\\
          \sig &  1.50\EE     &  1.41\EE     &  1.07\EE     &  9.38\E{- 1} &  7.03\E{- 1} &  7.48\E{- 1} &  5.93\E{- 1} &  6.27\E{- 1} \\
         \sige &  9.86\E{- 1} &  9.35\E{- 1} &  6.85\E{- 1} &  5.79\E{- 1} &  3.88\E{- 1} &  4.23\E{- 1} &  3.06\E{- 1} &  3.28\E{- 1} \\
         \sigi &  3.79\E{- 1} &  3.65\E{- 1} &  3.42\E{- 1} &  3.32\E{- 1} &  3.07\E{- 1} &  3.15\E{- 1} &  2.83\E{- 1} &  2.94\E{- 1} \\
         \sigr &  1.33\E{- 1} &  1.11\E{- 1} &  4.47\E{- 2} &  2.67\E{- 2} &  7.72\E{- 3} &  1.00\E{- 2} &  3.96\E{- 3} &  4.90\E{- 3} \\
         \sigp &  2.86\E{- 3} &  1.92\E{- 3} &  1.69\E{- 4} &  3.35\E{- 5} &  1.93\E{- 7} &  6.93\E{- 7} &  1.20\E{- 8} &  3.86\E{- 8} \\
&&&&&&&&&&&&\\
           \Yp &  6.80\E{- 9} &  1.42\E{- 7} &  2.93\E{- 9} &  1.97\E{- 9} &  1.54\E{-14} &  3.37\E{-12} &  4.02\E{- 8} &  7.28\E{- 7} \\
        \rtotp &  3.08\E{-11} &  1.26\E{- 9} &  1.79\E{-10} &  3.76\E{-10} &  5.16\E{-15} &  1.54\E{-12} &  2.40\E{- 6} &  6.72\E{- 5} \\
&&&&&&&&&&&&\\
        \wrate &  7.33\E{- 7} &  9.14\E{- 7} &  4.21\E{- 7} &  2.43\E{- 7} & -1.16\E{- 7} &  7.81\E{- 9} &  5.86\E{- 3} &  8.13\E{- 3} \\
\hline
\rate(\EC+\pd) &  \I{ 55}{Fe} &  \I{ 55}{Fe} &  \I{ 60}{Co} &  \I{ 57}{Fe} &  \I{ 60}{Co} &  \I{ 57}{Fe} &  \I{ 60}{Co} &  \I{ 65}{Ni} \\
               &  2.92\E{- 7} &  3.35\E{- 7} &  5.81\E{- 7} &  1.65\E{- 7} &  1.41\E{- 7} &  6.80\E{- 8} &  1.66\E{- 3} &  1.20\E{- 3} \\
               &  \I{ 53}{Mn} &  \I{ 54}{Fe} &  \I{ 59}{Co} &  \I{ 61}{Ni} &  \I{ 59}{Co} &  \I{ 55}{Mn} &  \I{ 58}{Mn} &  \I{ 59}{Fe} \\
               &  2.25\E{- 7} &  1.54\E{- 7} &  2.70\E{- 7} &  1.43\E{- 7} &  1.14\E{- 8} &  3.10\E{- 8} &  8.86\E{- 4} &  6.14\E{- 4} \\
               &  \I{ 57}{Co} &  \I{ 57}{Co} &  \I{ 54}{Mn} &  \I{ 56}{Fe} &  \I{ 57}{Fe} &  \I{ 61}{Ni} &  \I{ 59}{Fe} &  \I{ 52}{ V} \\
               &  5.46\E{- 8} &  9.59\E{- 8} &  2.51\E{- 7} &  8.81\E{- 8} &  7.16\E{- 9} &  2.63\E{- 8} &  8.27\E{- 4} &  5.91\E{- 4} \\
               &  \I{ 54}{Fe} &  \I{ 59}{Ni} &  \I{ 58}{Co} &  \I{ 55}{Mn} &  \I{ 55}{Mn} &  \I{ 53}{Cr} &  \I{ 52}{ V} &  \I{ 63}{Ni} \\
               &  4.24\E{- 8} &  8.75\E{- 8} &  1.81\E{- 7} &  6.54\E{- 8} &  4.32\E{- 9} &  1.63\E{- 8} &  7.50\E{- 4} &  5.70\E{- 4} \\
               &  \I{ 54}{Mn} &  \I{ 53}{Mn} &  \I{ 56}{Fe} &  \I{ 53}{Cr} &  \I{ 56}{Mn} &  \I{ 56}{Mn} &  \I{ 58}{Fe} &  \I{ 62}{Co} \\
               &  4.10\E{- 8} &  5.98\E{- 8} &  1.71\E{- 7} &  4.47\E{- 8} &  2.45\E{- 9} &  1.21\E{- 8} &  2.77\E{- 4} &  4.98\E{- 4} \\
&&&&&&&&&&&&\\
       \rectot &  7.52\E{- 7} &  9.09\E{- 7} &  1.94\E{- 6} &  7.10\E{- 7} &  1.69\E{- 7} &  2.04\E{- 7} &  5.87\E{- 3} &  8.14\E{- 3} \\
       \rpdtot &  9.49\E{-10} &  6.10\E{- 9} &  8.12\E{-11} &  5.44\E{-11} &  3.86\E{-15} &  4.88\E{-13} &  7.99\E{-11} &  5.03\E{- 9} \\
&&&&&&&&&&&&\\
\rate(\PC+\ed) &  \I{ 55}{Mn} &  \I{ 54}{Mn} &  \I{ 59}{Fe} &  \I{ 56}{Mn} &  \I{ 60}{Fe} &  \I{ 62}{Co} &  \I{ 58}{Mn} &  \I{ 64}{Co} \\
               &  1.22\E{- 8} &  2.42\E{-10} &  5.98\E{- 7} &  1.50\E{- 7} &  8.19\E{- 8} &  4.50\E{- 8} &  3.45\E{- 6} &  1.87\E{- 6} \\
               &  \I{ 53}{Cr} &  \I{ 58}{Co} &  \I{ 60}{Fe} &  \I{ 52}{ V} &  \I{ 57}{Mn} &  \I{ 58}{Mn} &  \I{ 54}{ V} &  \I{ 58}{Mn} \\
               &  2.73\E{- 9} &  1.44\E{-10} &  1.97\E{- 7} &  5.23\E{- 8} &  6.24\E{- 8} &  3.99\E{- 8} &  1.23\E{- 6} &  1.53\E{- 6} \\
               &  \I{ 54}{Cr} &  \I{ 56}{Mn} &  \I{ 57}{Mn} &  \I{ 57}{Mn} &  \I{ 59}{Fe} &  \I{ 56}{Mn} &  \I{ 56}{Cr} &  \I{ 54}{ V} \\
               &  2.62\E{- 9} &  1.26\E{-10} &  1.85\E{- 7} &  4.76\E{- 8} &  5.78\E{- 8} &  2.18\E{- 8} &  1.20\E{- 6} &  1.05\E{- 6} \\
               &  \I{ 57}{Fe} &  \I{ 55}{Mn} &  \I{ 54}{Cr} &  \I{ 62}{Co} &  \I{ 58}{Mn} &  \I{ 55}{Cr} &  \I{ 52}{Ti} &  \I{ 52}{Ti} \\
               &  1.10\E{- 9} &  1.19\E{-10} &  1.74\E{- 7} &  4.14\E{- 8} &  3.76\E{- 8} &  1.84\E{- 8} &  1.19\E{- 6} &  7.31\E{- 7} \\
               &  \I{ 56}{Mn} &  \I{ 60}{Co} &  \I{ 56}{Mn} &  \I{ 55}{Cr} &  \I{ 55}{Cr} &  \I{ 57}{Mn} &  \I{ 55}{Cr} &  \I{ 56}{Cr} \\
               &  8.99\E{-10} &  7.48\E{-11} &  8.48\E{- 8} &  4.05\E{- 8} &  2.94\E{- 8} &  1.81\E{- 8} &  3.54\E{- 7} &  5.57\E{- 7} \\
&&&&&&&&&&&&\\
       \redtot &  2.05\E{- 8} &  8.59\E{-10} &  1.52\E{- 6} &  4.67\E{- 7} &  2.86\E{- 7} &  1.96\E{- 7} &  7.93\E{- 6} &  8.08\E{- 6} \\
\hline
\end{tabular}
\end{table}

\clearpage

\begin{table}
\ifthenelse{\boolean{emul}}{}{\tiny}
\caption{Properties of the stellar center of a 25\,\Msun star \lTab{25}}
\begin{tabular}{l|r@{}lr@{}l|r@{}lr@{}l|r@{}lr@{}l|r@{}lr@{}l}
\hline
\hline
& \multicolumn{4}{c|}{O-dep}
& \multicolumn{4}{c|}{O-shell}
& \multicolumn{4}{c|}{Si-ign}
& \multicolumn{4}{c}{Si-burn} \\
& \multicolumn{2}{c}{FFN} & \multicolumn{2}{c|}{LMP}
& \multicolumn{2}{c}{FFN} & \multicolumn{2}{c|}{LMP}
& \multicolumn{2}{c}{FFN} & \multicolumn{2}{c|}{LMP}
& \multicolumn{2}{c}{FFN} & \multicolumn{2}{c}{LMP} \\
\hline
 $\log(\tb-t)$ &       5.90 & &       5.89 & &       5.40 & &       5.41 & &       4.79 & &       4.71 & &       4.40 & &       4.39 & \\
           $T$ &  2.49\E{  9} &  2.49\E{  9} &  2.36\E{  9} &  2.35\E{  9} &  3.21\E{  9} &  3.21\E{  9} &  3.62\E{  9} &  3.80\E{  9} \\
        $\rho$ &  5.73\E{  6} &  5.73\E{  6} &  2.61\E{  7} &  2.55\E{  7} &  6.88\E{  7} &  6.91\E{  7} &  2.17\E{  7} &  2.91\E{  7} \\
         $\Ye$ &  4.98\E{- 1} &  4.98\E{- 1} &  4.96\E{- 1} &  4.97\E{- 1} &  4.95\E{- 1} &  4.95\E{- 1} &  4.81\E{- 1} &  4.87\E{- 1} \\
&&&&&&&&&&&&\\
        \snucq &  1.79\E{ 12} &  1.79\E{ 12} &  7.60\E{  9} &  6.90\E{  9} &  1.42\E{ 13} &  1.36\E{ 13} &  1.04\E{ 14} &  5.50\E{ 13} \\
         \snuw &  1.99\E{  9} &  2.04\E{  9} &  3.73\E{  9} &  3.39\E{  9} &  4.16\E{ 11} &  1.88\E{ 11} &  6.11\E{ 12} &  8.06\E{ 12} \\
       \snubps &  1.34\E{ 12} &  1.34\E{ 12} &  8.02\E{ 10} &  8.10\E{ 10} &  6.63\E{ 11} &  6.59\E{ 11} &  1.34\E{ 13} &  1.52\E{ 13} \\
        \sdotq &  4.42\E{ 11} &  4.42\E{ 11} & -7.63\E{ 10} & -7.75\E{ 10} &  1.32\E{ 13} &  1.28\E{ 13} &  8.39\E{ 13} &  3.17\E{ 13} \\
&&&&&&&&&&&&\\
          \sig &  2.40\EE     &  2.40\EE     &  1.54\EE     &  1.54\EE     &  1.48\EE     &  1.48\EE     &  1.99\EE     &  1.91\EE     \\
         \sige &  1.42\EE     &  1.42\EE     &  8.39\E{- 1} &  8.44\E{- 1} &  8.03\E{- 1} &  8.02\E{- 1} &  1.25\EE     &  1.20\EE     \\
         \sigi &  6.44\E{- 1} &  6.44\E{- 1} &  6.38\E{- 1} &  6.39\E{- 1} &  6.18\E{- 1} &  6.18\E{- 1} &  4.64\E{- 1} &  4.76\E{- 1} \\
         \sigr &  3.28\E{- 1} &  3.28\E{- 1} &  6.11\E{- 2} &  6.21\E{- 2} &  5.85\E{- 2} &  5.83\E{- 2} &  2.65\E{- 1} &  2.28\E{- 1} \\
         \sigp &  1.14\E{- 2} &  1.14\E{- 2} &  1.64\E{- 4} &  1.71\E{- 4} &  2.48\E{- 4} &  2.45\E{- 4} &  1.38\E{- 2} &  1.03\E{- 2} \\
&&&&&&&&&&&&\\
           \Yp &  1.84\E{-11} &  1.84\E{-11} &  1.36\E{-13} &  1.34\E{-13} &  8.14\E{-10} &  8.99\E{-10} &  2.58\E{- 7} &  4.01\E{- 6} \\
        \rtotp &  1.40\E{-15} &  1.40\E{-15} &  9.21\E{-17} &  8.63\E{-17} &  7.41\E{-12} &  8.28\E{-12} &  4.74\E{-10} &  1.32\E{- 8} \\
&&&&&&&&&&&&\\
        \wrate &  1.43\E{- 9} &  1.41\E{- 9} &  3.14\E{- 9} &  2.84\E{- 9} &  2.55\E{- 7} &  1.14\E{- 7} &  3.53\E{- 6} &  3.45\E{- 6} \\
\hline
\rate(\EC+\pd) &  \I{ 37}{Ar} &  \I{ 37}{Ar} &  \I{ 35}{Cl} &  \I{ 35}{Cl} &  \I{ 55}{Fe} &  \I{ 35}{Cl} &  \I{ 54}{Fe} &  \I{ 53}{Fe} \\
               &  3.30\E{-10} &  3.41\E{-10} &  1.19\E{- 9} &  1.13\E{- 9} &  5.52\E{- 8} &  2.19\E{- 8} &  9.86\E{- 7} &  7.43\E{- 7} \\
               &  \I{ 32}{ S} &  \I{ 36}{Ar} &  \I{ 37}{Ar} &  \I{ 37}{Ar} &  \I{ 53}{Mn} &  \I{ 33}{ S} &  \I{ 55}{Co} &  \I{ 55}{Co} \\
               &  2.76\E{-10} &  2.92\E{-10} &  6.06\E{-10} &  5.73\E{-10} &  4.50\E{- 8} &  2.18\E{- 8} &  9.84\E{- 7} &  6.46\E{- 7} \\
               &  \I{ 35}{Cl} &  \I{ 35}{Cl} &  \I{ 33}{ S} &  \I{ 33}{ S} &  \I{ 54}{Fe} &  \I{ 32}{ S} &  \I{ 55}{Fe} &  \I{ 56}{Ni} \\
               &  2.70\E{-10} &  2.71\E{-10} &  5.73\E{-10} &  5.19\E{-10} &  3.20\E{- 8} &  1.92\E{- 8} &  4.88\E{- 7} &  6.10\E{- 7} \\
               &  \I{ 36}{Ar} &  \I{ 32}{ S} &  \I{ 32}{ S} &  \I{ 32}{ S} &  \I{ 33}{ S} &  \I{ 55}{Fe} &  \I{ 56}{Co} &  \I{ 54}{Fe} \\
               &  2.33\E{-10} &  2.54\E{-10} &  3.86\E{-10} &  3.33\E{-10} &  2.49\E{- 8} &  1.07\E{- 8} &  3.25\E{- 7} &  4.21\E{- 7} \\
               &  \I{ 45}{Ti} &  \I{ 45}{Ti} &  \I{ 36}{Ar} &  \I{ 36}{Ar} &  \I{ 35}{Cl} &  \I{ 37}{Ar} &  \I{ 53}{Mn} &  \I{ 57}{Ni} \\
               &  1.12\E{-10} &  1.12\E{-10} &  2.08\E{-10} &  2.02\E{-10} &  2.22\E{- 8} &  7.36\E{- 9} &  3.19\E{- 7} &  2.22\E{- 7} \\
&&&&&&&&&&&&\\
       \rectot &  1.40\E{- 9} &  1.38\E{- 9} &  3.13\E{- 9} &  2.83\E{- 9} &  2.54\E{- 7} &  1.13\E{- 7} &  3.49\E{- 6} &  3.30\E{- 6} \\
       \rpdtot &  3.04\E{-11} &  2.72\E{-11} &  7.92\E{-12} &  6.92\E{-12} &  1.08\E{- 9} &  9.40\E{-10} &  3.72\E{- 8} &  1.53\E{- 7} \\
&&&&&&&&&&&&\\
\rate(\PC+\ed) &  \I{ 32}{ P} &  \I{ 32}{ P} &  \I{ 32}{ P} &  \I{ 32}{ P} &  \I{ 32}{ P} &  \I{ 32}{ P} &  \I{ 55}{Mn} &  \I{ 54}{Mn} \\
               &  3.48\E{-14} &  1.86\E{-14} &  6.07\E{-14} &  3.13\E{-14} &  3.03\E{-11} &  1.41\E{-11} &  2.86\E{-10} &  8.01\E{-12} \\
               &  \I{ 36}{Cl} &  \I{ 36}{Cl} &  \I{ 36}{Cl} &  \I{ 36}{Cl} &  \I{ 55}{Mn} &  \I{ 54}{Mn} &  \I{ 53}{Cr} &  \I{ 55}{Fe} \\
               &  6.77\E{-15} &  5.98\E{-15} &  3.47\E{-15} &  3.01\E{-15} &  1.32\E{-11} &  1.82\E{-12} &  4.39\E{-11} &  2.75\E{-12} \\
               &  \I{ 46}{Sc} &  \I{ 46}{Sc} &  \I{ 46}{Sc} &  \I{ 54}{Mn} &  \I{ 53}{Cr} &  \I{ 36}{Cl} &  \I{ 54}{Mn} &  \I{ 58}{Co} \\
               &  2.60\E{-15} &  3.81\E{-16} &  1.05\E{-15} &  1.40\E{-16} &  2.66\E{-12} &  3.98\E{-13} &  3.29\E{-11} &  1.50\E{-12} \\
               &  \I{ 55}{Mn} &  \I{ 39}{Ar} &  \I{ 55}{Mn} &  \I{ 46}{Sc} &  \I{ 57}{Fe} &  \I{ 55}{Mn} &  \I{ 57}{Fe} &  \I{ 56}{Co} \\
               &  1.63\E{-16} &  1.54\E{-16} &  3.95\E{-16} &  1.31\E{-16} &  9.47\E{-13} &  3.13\E{-13} &  2.90\E{-11} &  1.26\E{-12} \\
               &  \I{ 33}{ P} &  \I{ 54}{Mn} &  \I{ 33}{ P} &  \I{ 39}{Ar} &  \I{ 54}{Mn} &  \I{ 28}{Al} &  \I{ 32}{ P} &  \I{ 32}{ P} \\
               &  1.55\E{-16} &  1.28\E{-16} &  3.30\E{-16} &  9.92\E{-17} &  6.09\E{-13} &  2.34\E{-13} &  1.17\E{-11} &  7.30\E{-13} \\
&&&&&&&&&&&&\\
       \redtot &  4.49\E{-14} &  2.54\E{-14} &  6.62\E{-14} &  3.48\E{-14} &  4.95\E{-11} &  1.77\E{-11} &  4.34\E{-10} &  1.62\E{-11} \\
\hline
\end{tabular}
\begin{flushright}\textsc{(continued on next page)}\end{flushright}
\end{table}

\clearpage

\addtocounter{table}{-1}

\begin{table}
\ifthenelse{\boolean{emul}}{}{\tiny}
\caption{(continued) Properties of the stellar center of a 25\,\Msun star}
\begin{tabular}{l|r@{}lr@{}l|r@{}lr@{}l|r@{}lr@{}l|r@{}lr@{}l}
\hline
\hline
& \multicolumn{4}{c|}{Si-dep}
& \multicolumn{4}{c|}{Si-shell}
& \multicolumn{4}{c|}{core-contr}
& \multicolumn{4}{c}{pre-SN} \\
& \multicolumn{2}{c}{FFN} & \multicolumn{2}{c|}{LMP}
& \multicolumn{2}{c}{FFN} & \multicolumn{2}{c|}{LMP}
& \multicolumn{2}{c}{FFN} & \multicolumn{2}{c|}{LMP}
& \multicolumn{2}{c}{FFN} & \multicolumn{2}{c}{LMP} \\
\hline
 $\log(\tb-t)$ &       4.05 & &       4.09 & &       3.67 & &       3.67 & &       3.27 & &       3.25 & &      -0.31 & &      -0.28 & \\
           $T$ &  3.89\E{  9} &  4.05\E{  9} &  4.81\E{  9} &  5.03\E{  9} &  4.44\E{  9} &  4.90\E{  9} &  7.13\E{  9} &  7.67\E{  9} \\
        $\rho$ &  2.52\E{  7} &  3.18\E{  7} &  1.37\E{  8} &  1.83\E{  8} &  2.26\E{  8} &  2.50\E{  8} &  2.65\E{  9} &  2.29\E{  9} \\
         $\Ye$ &  4.71\E{- 1} &  4.80\E{- 1} &  4.54\E{- 1} &  4.56\E{- 1} &  4.50\E{- 1} &  4.52\E{- 1} &  4.38\E{- 1} &  4.45\E{- 1} \\
&&&&&&&&&&&&\\
        \snucq &  3.10\E{ 13} &  3.09\E{ 13} & -5.24\E{ 11} & -3.71\E{ 11} & -3.13\E{ 11} & -3.40\E{ 11} & -8.71\E{ 15} & -1.49\E{ 15} \\
         \snuw &  1.22\E{ 13} &  1.24\E{ 13} &  2.70\E{ 13} &  1.00\E{ 13} &  1.78\E{ 13} &  8.54\E{ 12} &  1.15\E{ 16} &  5.54\E{ 15} \\
       \snubps &  2.36\E{ 13} &  2.66\E{ 13} &  2.15\E{ 13} &  2.23\E{ 13} &  3.76\E{ 12} &  1.00\E{ 13} &  6.99\E{ 12} &  2.49\E{ 13} \\
        \sdotq & -4.79\E{ 12} & -8.14\E{ 12} & -4.90\E{ 13} & -3.27\E{ 13} & -2.19\E{ 13} & -1.89\E{ 13} & -2.03\E{ 16} & -7.05\E{ 15} \\
&&&&&&&&&&&&\\
          \sig &  1.94\EE     &  1.88\EE     &  1.33\EE     &  1.27\EE     &  1.08\EE     &  1.14\EE     &  8.36\E{- 1} &  9.59\E{- 1} \\
         \sige &  1.25\EE     &  1.22\EE     &  8.72\E{- 1} &  8.34\E{- 1} &  6.90\E{- 1} &  7.36\E{- 1} &  4.86\E{- 1} &  5.51\E{- 1} \\
         \sigi &  3.88\E{- 1} &  3.88\E{- 1} &  3.56\E{- 1} &  3.55\E{- 1} &  3.42\E{- 1} &  3.45\E{- 1} &  3.34\E{- 1} &  3.84\E{- 1} \\
         \sigr &  2.84\E{- 1} &  2.54\E{- 1} &  9.83\E{- 2} &  8.45\E{- 2} &  4.67\E{- 2} &  5.73\E{- 2} &  1.66\E{- 2} &  2.39\E{- 2} \\
         \sigp &  1.79\E{- 2} &  1.43\E{- 2} &  1.78\E{- 3} &  1.22\E{- 3} &  2.13\E{- 4} &  4.12\E{- 4} &  8.57\E{- 6} &  3.19\E{- 5} \\
&&&&&&&&&&&&\\
           \Yp &  1.02\E{- 6} &  1.11\E{- 5} &  5.37\E{- 7} &  1.68\E{- 6} &  2.05\E{- 8} &  3.27\E{- 7} &  5.05\E{- 6} &  5.50\E{- 5} \\
        \rtotp &  2.84\E{- 9} &  4.93\E{- 8} &  2.94\E{- 8} &  1.59\E{- 7} &  2.27\E{- 9} &  4.95\E{- 8} &  5.68\E{- 5} &  5.15\E{- 4} \\
&&&&&&&&&&&&\\
        \wrate &  6.61\E{- 6} &  5.43\E{- 6} &  2.84\E{- 6} &  2.84\E{- 6} &  7.95\E{- 7} &  1.26\E{- 6} &  3.48\E{- 3} &  1.58\E{- 3} \\
\hline
\rate(\EC+\pd) &  \I{ 55}{Fe} &  \I{ 55}{Co} &  \I{ 60}{Co} &  \I{ 56}{Fe} &  \I{ 60}{Co} &  \I{ 56}{Fe} &  \I{ 60}{Co} &  \I{  1}{ H} \\
               &  1.43\E{- 6} &  1.02\E{- 6} &  1.85\E{- 6} &  6.68\E{- 7} &  2.84\E{- 6} &  4.52\E{- 7} &  2.28\E{- 3} &  5.15\E{- 4} \\
               &  \I{ 54}{Fe} &  \I{ 54}{Fe} &  \I{ 54}{Mn} &  \I{ 55}{Fe} &  \I{ 59}{Co} &  \I{ 57}{Fe} &  \I{ 57}{Fe} &  \I{ 53}{Cr} \\
               &  1.37\E{- 6} &  9.35\E{- 7} &  1.66\E{- 6} &  5.03\E{- 7} &  8.83\E{- 7} &  4.27\E{- 7} &  3.00\E{- 4} &  1.44\E{- 4} \\
               &  \I{ 55}{Co} &  \I{ 53}{Fe} &  \I{ 55}{Fe} &  \I{ 61}{Ni} &  \I{ 54}{Mn} &  \I{ 61}{Ni} &  \I{ 59}{Co} &  \I{ 57}{Fe} \\
               &  1.27\E{- 6} &  8.93\E{- 7} &  1.25\E{- 6} &  3.68\E{- 7} &  6.99\E{- 7} &  3.84\E{- 7} &  2.32\E{- 4} &  1.13\E{- 4} \\
               &  \I{ 56}{Co} &  \I{ 56}{Ni} &  \I{ 58}{Co} &  \I{ 54}{Mn} &  \I{ 57}{Fe} &  \I{ 54}{Mn} &  \I{ 52}{ V} &  \I{ 55}{Mn} \\
               &  8.43\E{- 7} &  6.77\E{- 7} &  9.95\E{- 7} &  3.49\E{- 7} &  4.16\E{- 7} &  2.28\E{- 7} &  1.88\E{- 4} &  9.11\E{- 5} \\
               &  \I{ 53}{Mn} &  \I{ 57}{Ni} &  \I{ 59}{Co} &  \I{ 57}{Fe} &  \I{ 56}{Fe} &  \I{ 55}{Mn} &  \I{ 54}{Mn} &  \I{ 56}{Fe} \\
               &  6.27\E{- 7} &  5.05\E{- 7} &  9.88\E{- 7} &  3.25\E{- 7} &  3.52\E{- 7} &  1.92\E{- 7} &  1.28\E{- 4} &  8.23\E{- 5} \\
&&&&&&&&&&&&\\
       \rectot &  6.56\E{- 6} &  5.20\E{- 6} &  9.54\E{- 6} &  3.85\E{- 6} &  6.39\E{- 6} &  2.84\E{- 6} &  3.76\E{- 3} &  1.63\E{- 3} \\
       \rpdtot &  5.78\E{- 8} &  2.34\E{- 7} &  1.21\E{- 9} &  5.74\E{- 9} &  1.23\E{-10} &  1.52\E{- 9} &  1.04\E{- 8} &  1.44\E{- 7} \\
&&&&&&&&&&&&\\
\rate(\PC+\ed) &  \I{ 55}{Mn} &  \I{ 54}{Mn} &  \I{ 59}{Fe} &  \I{ 56}{Mn} &  \I{ 59}{Fe} &  \I{ 56}{Mn} &  \I{ 58}{Mn} &  \I{ 58}{Mn} \\
               &  3.20\E{- 9} &  8.16\E{-11} &  1.85\E{- 6} &  4.52\E{- 7} &  1.97\E{- 6} &  5.83\E{- 7} &  6.60\E{- 5} &  9.19\E{- 6} \\
               &  \I{ 57}{Fe} &  \I{ 58}{Co} &  \I{ 57}{Mn} &  \I{ 52}{ V} &  \I{ 57}{Mn} &  \I{ 52}{ V} &  \I{ 56}{Cr} &  \I{ 55}{Cr} \\
               &  5.49\E{-10} &  3.17\E{-11} &  1.10\E{- 6} &  1.29\E{- 7} &  1.03\E{- 6} &  2.02\E{- 7} &  6.33\E{- 5} &  6.56\E{- 6} \\
               &  \I{ 53}{Cr} &  \I{ 55}{Fe} &  \I{ 54}{Cr} &  \I{ 60}{Co} &  \I{ 60}{Fe} &  \I{ 57}{Mn} &  \I{ 52}{Ti} &  \I{ 57}{Mn} \\
               &  4.55\E{-10} &  2.52\E{-11} &  1.03\E{- 6} &  9.57\E{- 8} &  9.00\E{- 7} &  1.29\E{- 7} &  5.48\E{- 5} &  3.72\E{- 6} \\
               &  \I{ 54}{Mn} &  \I{ 57}{Co} &  \I{ 60}{Fe} &  \I{ 57}{Mn} &  \I{ 54}{Cr} &  \I{ 55}{Cr} &  \I{ 55}{Cr} &  \I{ 56}{Cr} \\
               &  2.08\E{-10} &  8.73\E{-12} &  5.77\E{- 7} &  5.48\E{- 8} &  5.18\E{- 7} &  1.14\E{- 7} &  4.45\E{- 5} &  3.04\E{- 6} \\
               &  \I{ 58}{Co} &  \I{ 56}{Co} &  \I{ 55}{Mn} &  \I{ 55}{Cr} &  \I{ 55}{Cr} &  \I{ 60}{Co} &  \I{ 57}{Mn} &  \I{ 53}{ V} \\
               &  1.35\E{-10} &  6.96\E{-12} &  3.98\E{- 7} &  4.29\E{- 8} &  2.67\E{- 7} &  9.70\E{- 8} &  1.54\E{- 5} &  2.94\E{- 6} \\
&&&&&&&&&&&&\\
       \redtot &  4.84\E{- 9} &  1.73\E{-10} &  6.71\E{- 6} &  1.01\E{- 6} &  5.59\E{- 6} &  1.59\E{- 6} &  2.76\E{- 4} &  5.11\E{- 5} \\
\hline
\end{tabular}
\end{table}

\clearpage

\begin{table}
\ifthenelse{\boolean{emul}}{}{\tiny}
\caption{Properties of the stellar center of a 40\,\Msun star \lTab{40}}
\begin{tabular}{l|r@{}lr@{}l|r@{}lr@{}l|r@{}lr@{}l|r@{}lr@{}l}
\hline
\hline
& \multicolumn{4}{c|}{O-dep}
& \multicolumn{4}{c|}{O-shell}
& \multicolumn{4}{c|}{Si-ign}
& \multicolumn{4}{c}{Si-burn} \\
& \multicolumn{2}{c}{FFN} & \multicolumn{2}{c|}{LMP}
& \multicolumn{2}{c}{FFN} & \multicolumn{2}{c|}{LMP}
& \multicolumn{2}{c}{FFN} & \multicolumn{2}{c|}{LMP}
& \multicolumn{2}{c}{FFN} & \multicolumn{2}{c}{LMP} \\
\hline
 $\log(\tb-t)$ &       5.41 & &       5.41 & &       5.20 & &       5.20 & &       4.92 & &       4.84 & &       4.51 & &       4.53 & \\
           $T$ &  2.65\E{  9} &  2.65\E{  9} &  2.86\E{  9} &  2.85\E{  9} &  3.10\E{  9} &  3.13\E{  9} &  3.61\E{  9} &  3.70\E{  9} \\
        $\rho$ &  3.65\E{  6} &  3.65\E{  6} &  2.12\E{  7} &  2.06\E{  7} &  4.17\E{  7} &  4.59\E{  7} &  2.42\E{  7} &  3.61\E{  7} \\
         $\Ye$ &  4.98\E{- 1} &  4.98\E{- 1} &  4.96\E{- 1} &  4.97\E{- 1} &  4.94\E{- 1} &  4.96\E{- 1} &  4.87\E{- 1} &  4.93\E{- 1} \\
&&&&&&&&&&&&\\
        \snucq &  3.78\E{ 12} &  3.81\E{ 12} &  6.96\E{ 10} &  3.19\E{ 10} &  2.09\E{ 12} &  2.01\E{ 12} &  9.42\E{ 13} &  5.84\E{ 13} \\
         \snuw &  1.94\E{  9} &  1.93\E{  9} &  3.72\E{ 10} &  1.31\E{ 10} &  1.56\E{ 11} &  1.03\E{ 11} &  5.77\E{ 12} &  5.64\E{ 12} \\
       \snubps &  4.29\E{ 12} &  4.29\E{ 12} &  1.11\E{ 12} &  1.12\E{ 12} &  9.99\E{ 11} &  9.75\E{ 11} &  1.13\E{ 13} &  8.69\E{ 12} \\
        \sdotq & -5.10\E{ 11} & -4.78\E{ 11} & -1.07\E{ 12} & -1.10\E{ 12} &  9.31\E{ 11} &  9.31\E{ 11} &  7.71\E{ 13} &  4.41\E{ 13} \\
&&&&&&&&&&&&\\
          \sig &  3.03\EE     &  3.03\EE     &  1.85\EE     &  1.86\EE     &  1.63\EE     &  1.60\EE     &  1.99\EE     &  1.85\EE     \\
         \sige &  1.69\EE     &  1.69\EE     &  1.06\EE     &  1.07\EE     &  9.12\E{- 1} &  8.95\E{- 1} &  1.21\EE     &  1.11\EE     \\
         \sigi &  6.74\E{- 1} &  6.74\E{- 1} &  6.56\E{- 1} &  6.57\E{- 1} &  6.29\E{- 1} &  6.25\E{- 1} &  5.26\E{- 1} &  5.62\E{- 1} \\
         \sigr &  6.17\E{- 1} &  6.17\E{- 1} &  1.33\E{- 1} &  1.37\E{- 1} &  8.65\E{- 2} &  8.13\E{- 2} &  2.35\E{- 1} &  1.70\E{- 1} \\
         \sigp &  4.83\E{- 2} &  4.83\E{- 2} &  1.89\E{- 3} &  1.99\E{- 3} &  7.06\E{- 4} &  6.03\E{- 4} &  1.04\E{- 2} &  4.92\E{- 3} \\
&&&&&&&&&&&&\\
           \Yp &  1.96\E{-10} &  1.97\E{-10} &  9.42\E{-11} &  1.72\E{-10} &  3.11\E{-10} &  9.92\E{-10} &  2.63\E{- 7} &  2.13\E{- 6} \\
        \rtotp &  1.14\E{-14} &  1.14\E{-14} &  8.18\E{-14} &  1.41\E{-13} &  1.05\E{-12} &  4.11\E{-12} &  5.68\E{-10} &  9.07\E{- 9} \\
&&&&&&&&&&&&\\
        \wrate &  1.37\E{- 9} &  1.32\E{- 9} &  2.49\E{- 8} &  8.88\E{- 9} &  9.83\E{- 8} &  6.42\E{- 8} &  3.36\E{- 6} &  2.46\E{- 6} \\
\hline
\rate(\EC+\pd) &  \I{ 32}{ S} &  \I{ 37}{Ar} &  \I{ 54}{Fe} &  \I{ 55}{Fe} &  \I{ 53}{Mn} &  \I{ 55}{Fe} &  \I{ 55}{Co} &  \I{ 56}{Ni} \\
               &  2.88\E{-10} &  2.83\E{-10} &  6.38\E{- 9} &  1.72\E{- 9} &  2.81\E{- 8} &  1.27\E{- 8} &  1.15\E{- 6} &  5.32\E{- 7} \\
               &  \I{ 37}{Ar} &  \I{ 32}{ S} &  \I{ 53}{Mn} &  \I{ 32}{ S} &  \I{ 55}{Fe} &  \I{ 54}{Fe} &  \I{ 54}{Fe} &  \I{ 53}{Fe} \\
               &  2.74\E{-10} &  2.64\E{-10} &  5.63\E{- 9} &  1.37\E{- 9} &  2.39\E{- 8} &  1.17\E{- 8} &  9.18\E{- 7} &  4.90\E{- 7} \\
               &  \I{ 35}{Cl} &  \I{ 36}{Ar} &  \I{ 55}{Fe} &  \I{ 54}{Fe} &  \I{ 54}{Fe} &  \I{ 32}{ S} &  \I{ 55}{Fe} &  \I{ 55}{Co} \\
               &  2.41\E{-10} &  2.63\E{-10} &  4.98\E{- 9} &  1.17\E{- 9} &  1.48\E{- 8} &  6.62\E{- 9} &  3.59\E{- 7} &  4.42\E{- 7} \\
               &  \I{ 36}{Ar} &  \I{ 35}{Cl} &  \I{ 32}{ S} &  \I{ 35}{Cl} &  \I{ 33}{ S} &  \I{ 53}{Mn} &  \I{ 56}{Co} &  \I{ 54}{Fe} \\
               &  2.18\E{-10} &  2.40\E{-10} &  1.23\E{- 9} &  9.55\E{-10} &  6.06\E{- 9} &  6.22\E{- 9} &  2.92\E{- 7} &  2.51\E{- 7} \\
               &  \I{ 33}{ S} &  \I{ 33}{ S} &  \I{ 33}{ S} &  \I{ 33}{ S} &  \I{ 35}{Cl} &  \I{ 33}{ S} &  \I{ 53}{Mn} &  \I{ 52}{Fe} \\
               &  1.06\E{-10} &  9.51\E{-11} &  1.17\E{- 9} &  8.23\E{-10} &  4.23\E{- 9} &  4.73\E{- 9} &  2.44\E{- 7} &  1.48\E{- 7} \\
&&&&&&&&&&&&\\
       \rectot &  1.29\E{- 9} &  1.25\E{- 9} &  2.46\E{- 8} &  8.68\E{- 9} &  9.78\E{- 8} &  6.34\E{- 8} &  3.33\E{- 6} &  2.37\E{- 6} \\
       \rpdtot &  7.91\E{-11} &  7.04\E{-11} &  2.16\E{-10} &  2.07\E{-10} &  5.30\E{-10} &  7.85\E{-10} &  3.41\E{- 8} &  8.61\E{- 8} \\
&&&&&&&&&&&&\\
\rate(\PC+\ed) &  \I{ 32}{ P} &  \I{ 32}{ P} &  \I{ 32}{ P} &  \I{ 32}{ P} &  \I{ 32}{ P} &  \I{ 32}{ P} &  \I{ 55}{Mn} &  \I{ 54}{Mn} \\
               &  1.78\E{-13} &  9.10\E{-14} &  2.55\E{-12} &  5.22\E{-13} &  1.98\E{-11} &  3.32\E{-12} &  1.10\E{-10} &  1.63\E{-12} \\
               &  \I{ 36}{Cl} &  \I{ 36}{Cl} &  \I{ 55}{Mn} &  \I{ 54}{Mn} &  \I{ 55}{Mn} &  \I{ 54}{Mn} &  \I{ 53}{Cr} &  \I{ 55}{Fe} \\
               &  2.30\E{-14} &  2.16\E{-14} &  1.72\E{-12} &  2.33\E{-13} &  1.87\E{-11} &  2.24\E{-12} &  1.81\E{-11} &  5.77\E{-13} \\
               &  \I{ 46}{Sc} &  \I{ 46}{Sc} &  \I{ 53}{Cr} &  \I{ 36}{Cl} &  \I{ 53}{Cr} &  \I{ 55}{Mn} &  \I{ 54}{Mn} &  \I{ 32}{ P} \\
               &  7.10\E{-15} &  1.05\E{-15} &  2.20\E{-13} &  2.35\E{-14} &  3.88\E{-12} &  2.19\E{-13} &  1.63\E{-11} &  5.49\E{-13} \\
               &  \I{ 55}{Mn} &  \I{ 39}{Ar} &  \I{ 54}{Mn} &  \I{ 55}{Mn} &  \I{ 57}{Fe} &  \I{ 50}{ V} &  \I{ 32}{ P} &  \I{ 56}{Co} \\
               &  9.57\E{-16} &  6.59\E{-16} &  1.36\E{-13} &  1.79\E{-14} &  8.36\E{-13} &  1.55\E{-13} &  1.25\E{-11} &  3.14\E{-13} \\
               &  \I{ 33}{ P} &  \I{ 54}{Mn} &  \I{ 57}{Fe} &  \I{ 50}{ V} &  \I{ 54}{Mn} &  \I{ 53}{Cr} &  \I{ 57}{Fe} &  \I{ 58}{Co} \\
               &  8.98\E{-16} &  5.63\E{-16} &  6.56\E{-14} &  1.27\E{-14} &  7.45\E{-13} &  9.08\E{-14} &  1.08\E{-11} &  2.66\E{-13} \\
&&&&&&&&&&&&\\
       \redtot &  2.12\E{-13} &  1.16\E{-13} &  4.83\E{-12} &  8.38\E{-13} &  4.60\E{-11} &  6.43\E{-12} &  1.82\E{-10} &  3.75\E{-12} \\
\hline
\end{tabular}
\begin{flushright}\textsc{(continued on next page)}\end{flushright}
\end{table}

\clearpage

\addtocounter{table}{-1}

\begin{table}
\ifthenelse{\boolean{emul}}{}{\tiny}
\caption{(continued) Properties of the stellar center of a 40\,\Msun star}
\begin{tabular}{l|r@{}lr@{}l|r@{}lr@{}l|r@{}lr@{}l|r@{}lr@{}l}
\hline
\hline
& \multicolumn{4}{c|}{Si-dep}
& \multicolumn{4}{c|}{Si-shell}
& \multicolumn{4}{c|}{core-contr}
& \multicolumn{4}{c}{pre-SN} \\
& \multicolumn{2}{c}{FFN} & \multicolumn{2}{c|}{LMP}
& \multicolumn{2}{c}{FFN} & \multicolumn{2}{c|}{LMP}
& \multicolumn{2}{c}{FFN} & \multicolumn{2}{c|}{LMP}
& \multicolumn{2}{c}{FFN} & \multicolumn{2}{c}{LMP} \\
\hline
 $\log(\tb-t)$ &       3.78 & &       3.75 & &       3.11 & &       3.15 & &       2.65 & &       2.63 & &      -0.33 & &      -0.22 & \\
           $T$ &  3.98\E{  9} &  4.11\E{  9} &  5.17\E{  9} &  5.38\E{  9} &  5.17\E{  9} &  5.62\E{  9} &  7.68\E{  9} &  8.01\E{  9} \\
        $\rho$ &  2.47\E{  7} &  2.98\E{  7} &  1.35\E{  8} &  1.47\E{  8} &  2.47\E{  8} &  2.47\E{  8} &  2.50\E{  9} &  1.68\E{  9} \\
         $\Ye$ &  4.71\E{- 1} &  4.80\E{- 1} &  4.54\E{- 1} &  4.59\E{- 1} &  4.49\E{- 1} &  4.53\E{- 1} &  4.39\E{- 1} &  4.47\E{- 1} \\
&&&&&&&&&&&&\\
        \snucq &  2.50\E{ 13} &  3.32\E{ 13} & -8.67\E{ 11} &  2.16\E{ 12} & -1.35\E{ 12} & -6.10\E{ 11} & -1.10\E{ 16} &  5.76\E{ 14} \\
         \snuw &  1.62\E{ 13} &  1.46\E{ 13} &  5.26\E{ 13} &  2.00\E{ 13} &  8.26\E{ 13} &  3.09\E{ 13} &  1.62\E{ 16} &  6.71\E{ 15} \\
       \snubps &  3.05\E{ 13} &  3.30\E{ 13} &  4.72\E{ 13} &  6.30\E{ 13} &  1.88\E{ 13} &  4.63\E{ 13} &  2.12\E{ 13} &  8.30\E{ 13} \\
        \sdotq & -2.17\E{ 13} & -1.44\E{ 13} & -1.01\E{ 14} & -8.08\E{ 13} & -1.03\E{ 14} & -7.79\E{ 13} & -2.72\E{ 16} & -6.22\E{ 15} \\
&&&&&&&&&&&&\\
          \sig &  2.00\EE     &  1.95\EE     &  1.43\EE     &  1.46\EE     &  1.19\EE     &  1.29\EE     &  9.26\E{- 1} &  1.15\EE     \\
         \sige &  1.28\EE     &  1.26\EE     &  9.34\E{- 1} &  9.50\E{- 1} &  7.71\E{- 1} &  8.35\E{- 1} &  5.31\E{- 1} &  6.33\E{- 1} \\
         \sigi &  3.84\E{- 1} &  3.89\E{- 1} &  3.64\E{- 1} &  3.72\E{- 1} &  3.49\E{- 1} &  3.62\E{- 1} &  3.72\E{- 1} &  4.77\E{- 1} \\
         \sigr &  3.11\E{- 1} &  2.83\E{- 1} &  1.24\E{- 1} &  1.29\E{- 1} &  6.81\E{- 2} &  8.73\E{- 2} &  2.21\E{- 2} &  3.72\E{- 2} \\
         \sigp &  2.23\E{- 2} &  1.83\E{- 2} &  3.40\E{- 3} &  3.73\E{- 3} &  7.04\E{- 4} &  1.44\E{- 3} &  2.48\E{- 5} &  1.39\E{- 4} \\
&&&&&&&&&&&&\\
           \Yp &  2.66\E{- 6} &  1.84\E{- 5} &  2.86\E{- 6} &  1.47\E{- 5} &  6.06\E{- 7} &  6.90\E{- 6} &  2.09\E{- 5} &  1.77\E{- 4} \\
        \rtotp &  7.71\E{- 9} &  7.76\E{- 8} &  1.75\E{- 7} &  1.12\E{- 6} &  9.52\E{- 8} &  1.24\E{- 6} &  2.21\E{- 4} &  1.03\E{- 3} \\
&&&&&&&&&&&&\\
        \wrate &  8.64\E{- 6} &  6.29\E{- 6} &  6.36\E{- 6} &  6.84\E{- 6} &  2.21\E{- 6} &  4.69\E{- 6} &  4.37\E{- 3} &  1.76\E{- 3} \\
\hline
\rate(\EC+\pd) &  \I{ 55}{Co} &  \I{ 55}{Co} &  \I{ 54}{Mn} &  \I{ 55}{Fe} &  \I{ 60}{Co} &  \I{ 56}{Fe} &  \I{ 60}{Co} &  \I{  1}{ H} \\
               &  1.83\E{- 6} &  1.21\E{- 6} &  3.37\E{- 6} &  1.37\E{- 6} &  1.17\E{- 5} &  1.27\E{- 6} &  2.73\E{- 3} &  1.03\E{- 3} \\
               &  \I{ 55}{Fe} &  \I{ 53}{Fe} &  \I{ 60}{Co} &  \I{  1}{ H} &  \I{ 59}{Co} &  \I{  1}{ H} &  \I{ 57}{Fe} &  \I{ 53}{Cr} \\
               &  1.60\E{- 6} &  1.04\E{- 6} &  3.25\E{- 6} &  1.12\E{- 6} &  3.21\E{- 6} &  1.24\E{- 6} &  3.76\E{- 4} &  9.73\E{- 5} \\
               &  \I{ 56}{Co} &  \I{ 54}{Fe} &  \I{ 55}{Fe} &  \I{ 56}{Fe} &  \I{ 54}{Mn} &  \I{ 57}{Fe} &  \I{ 59}{Co} &  \I{ 57}{Fe} \\
               &  1.45\E{- 6} &  9.44\E{- 7} &  2.55\E{- 6} &  1.05\E{- 6} &  3.17\E{- 6} &  9.44\E{- 7} &  3.02\E{- 4} &  7.53\E{- 5} \\
               &  \I{ 54}{Fe} &  \I{ 56}{Ni} &  \I{ 58}{Co} &  \I{ 54}{Mn} &  \I{ 57}{Fe} &  \I{ 54}{Mn} &  \I{ 52}{ V} &  \I{ 56}{Fe} \\
               &  1.27\E{- 6} &  9.15\E{- 7} &  1.79\E{- 6} &  5.93\E{- 7} &  1.69\E{- 6} &  7.83\E{- 7} &  2.61\E{- 4} &  6.35\E{- 5} \\
               &  \I{ 57}{Co} &  \I{ 57}{Ni} &  \I{ 59}{Co} &  \I{ 53}{Mn} &  \I{ 58}{Co} &  \I{ 61}{Ni} &  \I{ 54}{Mn} &  \I{ 55}{Mn} \\
               &  8.86\E{- 7} &  6.30\E{- 7} &  1.76\E{- 6} &  4.90\E{- 7} &  1.17\E{- 6} &  7.56\E{- 7} &  2.24\E{- 4} &  6.32\E{- 5} \\
&&&&&&&&&&&&\\
       \rectot &  8.57\E{- 6} &  5.99\E{- 6} &  1.82\E{- 5} &  7.86\E{- 6} &  2.54\E{- 5} &  9.31\E{- 6} &  4.90\E{- 3} &  1.84\E{- 3} \\
       \rpdtot &  8.55\E{- 8} &  3.00\E{- 7} &  2.90\E{- 9} &  2.98\E{- 8} &  5.41\E{-10} &  1.22\E{- 8} &  4.57\E{- 8} &  3.03\E{- 7} \\
&&&&&&&&&&&&\\
\rate(\PC+\ed) &  \I{ 55}{Mn} &  \I{ 54}{Mn} &  \I{ 59}{Fe} &  \I{ 56}{Mn} &  \I{ 57}{Mn} &  \I{ 56}{Mn} &  \I{ 56}{Cr} &  \I{ 58}{Mn} \\
               &  3.43\E{- 9} &  7.95\E{-11} &  2.82\E{- 6} &  4.29\E{- 7} &  6.49\E{- 6} &  1.48\E{- 6} &  1.26\E{- 4} &  1.22\E{- 5} \\
               &  \I{ 57}{Fe} &  \I{ 58}{Co} &  \I{ 57}{Mn} &  \I{ 52}{ V} &  \I{ 59}{Fe} &  \I{ 52}{ V} &  \I{ 58}{Mn} &  \I{ 55}{Cr} \\
               &  1.06\E{- 9} &  3.42\E{-11} &  2.26\E{- 6} &  1.13\E{- 7} &  5.88\E{- 6} &  5.54\E{- 7} &  1.17\E{- 4} &  9.82\E{- 6} \\
               &  \I{ 58}{Co} &  \I{ 55}{Fe} &  \I{ 54}{Cr} &  \I{ 60}{Co} &  \I{ 60}{Fe} &  \I{ 57}{Mn} &  \I{ 52}{Ti} &  \I{ 57}{Mn} \\
               &  3.99\E{-10} &  2.98\E{-11} &  1.87\E{- 6} &  1.09\E{- 7} &  3.16\E{- 6} &  3.33\E{- 7} &  1.02\E{- 4} &  5.91\E{- 6} \\
               &  \I{ 53}{Cr} &  \I{ 57}{Co} &  \I{ 60}{Fe} &  \I{ 55}{Mn} &  \I{ 54}{Cr} &  \I{ 55}{Cr} &  \I{ 55}{Cr} &  \I{ 56}{Mn} \\
               &  3.07\E{-10} &  1.09\E{-11} &  8.87\E{- 7} &  7.39\E{- 8} &  1.89\E{- 6} &  3.23\E{- 7} &  8.57\E{- 5} &  5.34\E{- 6} \\
               &  \I{ 54}{Mn} &  \I{ 56}{Co} &  \I{ 55}{Mn} &  \I{ 53}{Cr} &  \I{ 55}{Cr} &  \I{ 60}{Co} &  \I{ 57}{Mn} &  \I{ 53}{ V} \\
               &  1.99\E{-10} &  9.52\E{-12} &  7.09\E{- 7} &  4.04\E{- 8} &  1.63\E{- 6} &  2.61\E{- 7} &  2.96\E{- 5} &  4.48\E{- 6} \\
&&&&&&&&&&&&\\
       \redtot &  5.86\E{- 9} &  1.83\E{-10} &  1.18\E{- 5} &  1.04\E{- 6} &  2.32\E{- 5} &  4.63\E{- 6} &  5.28\E{- 4} &  7.76\E{- 5} \\
\hline
\end{tabular}
\end{table}
}

\clearpage

{

\renewcommand{\E}[1]{&{\ensuremath{\cdot10^{#1}}}}
\newcommand{\EE}{&}
\renewcommand{\I}[2]{{\ensuremath{^{#1}}}&{\ensuremath{\mathrm{#2}}}}
\newcommand{\NoData}{\multicolumn{2}{c}{\nodata}}

\begin{table}
{\centering
\caption{$\int$ rate / total rate d$\Ye$ -- most important contributors to $\Delta\Ye$, LMP $15\,\Msun$ \lTab{flow15}}
\begin{tabular}{r@{}lr@{}lr@{}lr@{}lr@{}lr@{}lr@{}lr@{}lr@{}lr@{}l}
\hline
\hline
\multicolumn{2}{c}{ion} &
\multicolumn{2}{c}{total} &
\multicolumn{2}{c}{\EC} &
\multicolumn{2}{c}{\ed} &
\multicolumn{2}{c}{\pd} \\
\hline
 \I{ 57}{Fe} &  1.13\E{- 2} &  1.13\E{- 2} &  4.64\E{- 5} &  4.25\E{- 8} \\
 \I{ 55}{Fe} &  1.01\E{- 2} &  1.00\E{- 2} &  8.33\E{- 8} &  9.91\E{- 6} \\
 \I{ 61}{Ni} &  7.56\E{- 3} &  7.56\E{- 3} &  1.47\E{- 7} &  7.44\E{- 8} \\
 \I{ 54}{Fe} &  6.92\E{- 3} &  6.91\E{- 3} &  4.17\E{-13} &  9.52\E{- 6} \\
 \I{ 56}{Fe} &  5.87\E{- 3} &  5.87\E{- 3} &  4.78\E{- 7} &  1.43\E{- 7} \\
 \I{ 55}{Mn} &  5.15\E{- 3} &  5.28\E{- 3} &  1.24\E{- 4} &  4.38\E{- 8} \\
 \I{ 53}{Cr} &  4.28\E{- 3} &  4.33\E{- 3} &  5.27\E{- 5} &  9.19\E{- 9} \\
 \I{ 35}{Cl} &  3.48\E{- 3} &  3.48\E{- 3} &  9.81\E{-15} &  2.86\E{- 8} \\
 \I{ 53}{Mn} &  3.44\E{- 3} &  3.44\E{- 3} &  5.53\E{- 9} &  7.34\E{- 6} \\
 \I{ 33}{ S} &  3.09\E{- 3} &  3.09\E{- 3} &  3.36\E{-14} &  4.19\E{- 9} \\
\hline
 \I{ 56}{Mn} & -4.05\E{- 3} &  2.23\E{- 3} &  6.28\E{- 3} &  2.89\E{- 9} \\
 \I{ 58}{Mn} & -4.32\E{- 3} &  4.34\E{- 5} &  4.37\E{- 3} &  8.64\E{-12} \\
 \I{ 62}{Co} & -3.54\E{- 3} &  2.91\E{- 4} &  3.83\E{- 3} &  7.06\E{-11} \\
 \I{ 57}{Mn} & -2.34\E{- 3} &  2.36\E{- 5} &  2.36\E{- 3} &  2.47\E{-11} \\
 \I{ 55}{Cr} & -2.29\E{- 3} &  4.10\E{- 5} &  2.33\E{- 3} &  2.30\E{-11} \\
 \I{ 52}{ V} & -5.33\E{- 4} &  1.61\E{- 3} &  2.14\E{- 3} &  7.96\E{-10} \\
 \I{ 53}{ V} & -1.54\E{- 3} &  8.92\E{- 5} &  1.63\E{- 3} &  2.18\E{-11} \\
 \I{ 63}{Co} & -9.55\E{- 4} &  2.56\E{- 5} &  9.81\E{- 4} &  1.53\E{-12} \\
 \I{ 60}{Co} &  8.18\E{- 5} &  1.06\E{- 3} &  9.76\E{- 4} &  5.45\E{- 8} \\
 \I{ 64}{Co} & -7.20\E{- 4} &      \NoData &  7.20\E{- 4} &      \NoData \\
\hline
\end{tabular}
\\[0.5\baselineskip] 
}
\textsc{Note.}---
The first section list the ten ions most important for decreasing \Ye
(electron capture and $\beta^+$-decay), the second section those which
dominate $\beta^-$-decay.
\end{table}

\clearpage

\begin{table}
\centering
\caption{$\int$ rate / total rate d$\Ye$ -- most important contributors to $\Delta\Ye$, LMP $25\,\Msun$ \lTab{flow25}}
\begin{tabular}{r@{}lr@{}lr@{}lr@{}lr@{}lr@{}lr@{}lr@{}lr@{}lr@{}l}
\hline
\hline
\multicolumn{2}{c}{ion} &
\multicolumn{2}{c}{total} &
\multicolumn{2}{c}{\EC} &
\multicolumn{2}{c}{\ed} &
\multicolumn{2}{c}{\pd} \\
\hline
 \I{ 55}{Fe} &  6.27\E{- 3} &  6.25\E{- 3} &  5.36\E{- 7} &  1.93\E{- 5} \\
 \I{ 54}{Fe} &  5.00\E{- 3} &  4.97\E{- 3} &  2.54\E{-11} &  2.53\E{- 5} \\
 \I{ 56}{Fe} &  4.45\E{- 3} &  4.45\E{- 3} &  2.22\E{- 6} &  6.32\E{- 7} \\
 \I{ 55}{Co} &  4.36\E{- 3} &  4.26\E{- 3} &  7.73\E{-15} &  1.03\E{- 4} \\
 \I{ 53}{Fe} &  4.23\E{- 3} &  3.87\E{- 3} &  6.05\E{-16} &  3.57\E{- 4} \\
 \I{ 56}{Ni} &  3.69\E{- 3} &  3.68\E{- 3} &      \NoData &  1.27\E{- 5} \\
 \I{ 57}{Fe} &  3.44\E{- 3} &  3.52\E{- 3} &  8.11\E{- 5} &  1.89\E{- 7} \\
 \I{ 61}{Ni} &  2.90\E{- 3} &  2.90\E{- 3} &  6.16\E{- 7} &  2.73\E{- 7} \\
 \I{ 54}{Mn} &  2.28\E{- 3} &  2.30\E{- 3} &  2.95\E{- 5} &  8.45\E{- 6} \\
 \I{ 57}{Ni} &  1.90\E{- 3} &  1.86\E{- 3} &  3.13\E{-14} &  4.26\E{- 5} \\
\hline
 \I{ 56}{Mn} & -3.27\E{- 3} &  4.42\E{- 4} &  3.71\E{- 3} &  1.65\E{- 8} \\
 \I{ 52}{ V} & -1.01\E{- 3} &  2.69\E{- 4} &  1.28\E{- 3} &  7.34\E{- 9} \\
 \I{ 58}{Mn} & -9.74\E{- 4} &  4.57\E{- 7} &  9.74\E{- 4} &  2.28\E{-11} \\
 \I{ 55}{Cr} & -9.38\E{- 4} &  1.42\E{- 6} &  9.39\E{- 4} &  1.91\E{-10} \\
 \I{ 57}{Mn} & -9.22\E{- 4} &  9.05\E{- 7} &  9.23\E{- 4} &  2.35\E{-10} \\
 \I{ 62}{Co} & -6.59\E{- 4} &  8.46\E{- 6} &  6.68\E{- 4} &  2.31\E{-10} \\
 \I{ 60}{Co} & -2.78\E{- 4} &  3.76\E{- 4} &  6.55\E{- 4} &  1.67\E{- 7} \\
 \I{ 53}{ V} & -6.05\E{- 4} &  2.65\E{- 6} &  6.08\E{- 4} &  1.32\E{-10} \\
 \I{ 59}{Fe} & -4.07\E{- 4} &  2.36\E{- 5} &  4.30\E{- 4} &  9.14\E{-10} \\
 \I{ 61}{Co} & -2.78\E{- 4} &  3.50\E{- 5} &  3.13\E{- 4} &  3.03\E{- 9} \\
\hline
\end{tabular}
\end{table}

\clearpage

\begin{table}
\centering
\caption{$\int$ rate / total rate d$\Ye$ -- most important contributors to $\Delta\Ye$, LMP $40\,\Msun$ \lTab{flow40}}
\begin{tabular}{r@{}lr@{}lr@{}lr@{}lr@{}lr@{}lr@{}lr@{}lr@{}lr@{}l}
\hline
\hline
\multicolumn{2}{c}{ion} &
\multicolumn{2}{c}{total} &
\multicolumn{2}{c}{\EC} &
\multicolumn{2}{c}{\ed} &
\multicolumn{2}{c}{\pd} \\
\hline
 \I{ 55}{Fe} &  5.93\E{- 3} &  5.91\E{- 3} &  6.91\E{- 7} &  2.27\E{- 5} \\
 \I{ 54}{Fe} &  5.21\E{- 3} &  5.19\E{- 3} &  4.40\E{-11} &  2.77\E{- 5} \\
 \I{  1}{ H} &  5.01\E{- 3} &  5.01\E{- 3} &      \NoData &      \NoData \\
 \I{ 55}{Co} &  4.38\E{- 3} &  4.27\E{- 3} &  1.35\E{-14} &  1.15\E{- 4} \\
 \I{ 53}{Fe} &  4.28\E{- 3} &  3.90\E{- 3} &  1.03\E{-15} &  3.79\E{- 4} \\
 \I{ 56}{Ni} &  3.74\E{- 3} &  3.72\E{- 3} &      \NoData &  1.57\E{- 5} \\
 \I{ 56}{Fe} &  3.04\E{- 3} &  3.04\E{- 3} &  3.90\E{- 6} &  1.26\E{- 6} \\
 \I{ 57}{Ni} &  1.95\E{- 3} &  1.90\E{- 3} &  5.62\E{-14} &  4.77\E{- 5} \\
 \I{ 54}{Mn} &  1.80\E{- 3} &  1.82\E{- 3} &  3.12\E{- 5} &  1.04\E{- 5} \\
 \I{ 53}{Mn} &  1.80\E{- 3} &  1.78\E{- 3} &  8.13\E{- 8} &  1.17\E{- 5} \\
\hline
 \I{ 56}{Mn} & -1.93\E{- 3} &  2.10\E{- 4} &  2.14\E{- 3} &  3.21\E{- 8} \\
 \I{ 52}{ V} & -6.49\E{- 4} &  1.32\E{- 4} &  7.80\E{- 4} &  1.66\E{- 8} \\
 \I{ 55}{Cr} & -5.32\E{- 4} &  8.86\E{- 7} &  5.33\E{- 4} &  5.00\E{-10} \\
 \I{ 57}{Mn} & -5.03\E{- 4} &  5.89\E{- 7} &  5.03\E{- 4} &  5.84\E{-10} \\
 \I{ 58}{Mn} & -4.70\E{- 4} &  2.39\E{- 7} &  4.70\E{- 4} &  4.78\E{-11} \\
 \I{ 60}{Co} & -2.01\E{- 4} &  2.06\E{- 4} &  4.07\E{- 4} &  2.39\E{- 7} \\
 \I{ 53}{ V} & -3.30\E{- 4} &  1.58\E{- 6} &  3.32\E{- 4} &  3.31\E{-10} \\
 \I{ 59}{Fe} & -2.93\E{- 4} &  1.11\E{- 5} &  3.05\E{- 4} &  2.09\E{- 9} \\
 \I{ 62}{Co} & -2.96\E{- 4} &  3.42\E{- 6} &  2.99\E{- 4} &  3.89\E{-10} \\
 \I{ 54}{Cr} & -2.44\E{- 4} &  2.73\E{- 5} &  2.71\E{- 4} &  7.43\E{- 9} \\
\hline
\end{tabular}
\end{table}

\clearpage

\begin{table}
\ifthenelse{\boolean{emul}}{}{\tiny}
{\centering
\caption{Integrated values, summary table \lTab{flow-sum}}
\begin{tabular}{r@{}lr@{}lr@{}lr@{}lr@{}lr@{}lr@{}lr@{}lr@{}lr@{}lr@{}lr@{}lr@{}lr@{}lr@{}lr@{}lr@{}lr@{}l}
\hline
\hline
\multicolumn{2}{c}{model} &
\multicolumn{2}{c}{\F } &
\multicolumn{2}{c}{\FEC} &
\multicolumn{2}{c}{\Fpd} &
\multicolumn{2}{c}{\Fed} &
\multicolumn{2}{c}{\Eweak} &
\multicolumn{2}{c}{\Eplas} &
\multicolumn{2}{c}{\EEC} &
\multicolumn{2}{c}{\Eed} \\
\multicolumn{2}{c}{  } &
\multicolumn{2}{c}{  } &
\multicolumn{2}{c}{  } &
\multicolumn{2}{c}{  } &
\multicolumn{2}{c}{  } &
\multicolumn{2}{c}{\ergg } &
\multicolumn{2}{c}{\ergg } &
\multicolumn{2}{c}{\ergg } &
\multicolumn{2}{c}{\ergg } \\
\hline
FFN &, $15\,\Msun$ & 3.39\E{- 1} &  3.94\E{- 1} &  7.04\E{- 4} &  5.49\E{- 2} &  7.09\E{ 17} &  1.20\E{ 18} &  6.34\E{ 17} &  7.44\E{ 16} \\
LMP &, $15\,\Msun$ & 2.60\E{- 1} &  2.86\E{- 1} &  1.98\E{- 3} &  2.88\E{- 2} &  5.39\E{ 17} &  1.24\E{ 18} &  4.88\E{ 17} &  5.10\E{ 16} \\
FFN &, $25\,\Msun$ & 2.30\E{- 1} &  2.78\E{- 1} &  1.84\E{- 3} &  4.96\E{- 2} &  5.60\E{ 17} &  1.00\E{ 18} &  4.82\E{ 17} &  7.76\E{ 16} \\
LMP &, $25\,\Msun$ & 1.63\E{- 1} &  1.70\E{- 1} &  5.49\E{- 3} &  1.19\E{- 2} &  4.16\E{ 17} &  9.94\E{ 17} &  3.92\E{ 17} &  2.40\E{ 16} \\
FFN &, $40\,\Msun$ & 2.34\E{- 1} &  2.80\E{- 1} &  2.14\E{- 3} &  4.76\E{- 2} &  5.82\E{ 17} &  1.27\E{ 18} &  4.99\E{ 17} &  8.34\E{ 16} \\
LMP &, $40\,\Msun$ & 1.71\E{- 1} &  1.72\E{- 1} &  6.44\E{- 3} &  7.23\E{- 3} &  4.25\E{ 17} &  1.26\E{ 18} &  4.10\E{ 17} &  1.58\E{ 16} \\
\hline
\end{tabular}
\\[0.5\baselineskip] 
}
\textsc{Note.}---
\F is the time-integrated central weak flow
($F_i\equiv\int_{t_\mathrm{O-dep}}^{t_\mathrm{pre-SN}}
\rate_{i,\mathrm{central}}(t)\;\mathrm{d}t$), and respectively \FEC,
\Fpd, and \Fed are the contributions due to electron capture,
\ed-decay and \ed-decay, $\F=\FEC+\Fpd-\Fed$. \Eweak, \Eplas, \EEC,
and \Eed are the total specific energy loss by neutrinos due to all
weak interactions, plasma neutrinos, electron capture and \ed-decay.
\end{table}

}


\ifthenelse{\boolean{emul}}{}{
\clearpage
\onecolumn

\newcommand{\FC}[3]{\figcaption[#1]{#3}}
\newcommand{\FCff}[3][]{}

\ifthenelse{\boolean{\IncludeFigures}}{
\renewcommand{\FC}[3][1.0]{
\clearpage
\begin{figure}
\epsscale{#1}
\plotone{#2}
\caption{#3}
\end{figure}
}
\renewcommand{\FCff}[3][1.0]{
\addtocounter{figure}{-1}
\FC[#1]{#2}{#3}
}}{}

\FC[0.6]{\FigOneFile}{\FigOne}
\FC{\FigTwoAFile}{\FigTwo}
\FCff{\FigTwoBFile}{(Fig.~2b)}
\FC{\FigThreeFile}{\FigThree}
\FC{\FigFourFile}{\FigFour}
\FC{\FigFiveFile}{\FigFive}
\FC{\FigSixFile}{\FigSix}
\FC{\FigSevenFile}{\FigSeven}
\FC{\FigEightFile}{\FigEight}
\FC{\FigNineFile}{\FigNine}
\FC{\FigTenFile}{\FigTen}
\FC[0.5]{\FigElevenFile}{\FigEleven}
\FC[0.5]{\FigTwelveFile}{\FigTwelve}
\FC[0.7]{\FigThirteenFile}{\FigThirteen}
\FC{\FigFourteenFile}{\FigFourteen}
\FC{\FigFifteenFile}{\FigFifteen}
\FC{\FigSixteenFile}{\FigSixteen}
}

\end{document}